\documentclass[%
aip,
 amsmath,amssymb,
reprint,%
]{revtex4-1}

\usepackage{graphicx}
\usepackage{dcolumn}
\usepackage{bm}

\usepackage[utf8]{inputenc}
\usepackage[T1]{fontenc}
\usepackage{mathptmx}
\usepackage{etoolbox}

\makeatletter
\def\@email#1#2{%
 \endgroup
 \patchcmd{\titleblock@produce}
  {\frontmatter@RRAPformat}
  {\frontmatter@RRAPformat{\produce@RRAP{*#1\href{mailto:#2}{#2}}}\frontmatter@RRAPformat}
  {}{}
}%
\makeatother
\begin{document}

\preprint{AIP/123-QED}

\title[Mixed discrete variable Gaussian states]{Mixed discrete variable Gaussian states}
\author{Nicolae Cotfas}
 \email{ncotfas@yahoo.com \ ,\ nicolae.cotfas@unibuc.ro , https://unibuc.ro/user/nicolae.cotfas/}
\affiliation{ 
Physics Department, University of Bucharest
}%

\date{\today}
\begin{abstract}
The quantum systems with finite-dimensional Hilbert space have several applications and are intensively explored theoretically and experimentally. 
The mathematical description of these systems follows the analogy with the usual infinite-dimensional case.
There exist finite versions for most of the elements used in the continuous case, but (to our knowledge) there does not exist a finite version corresponding to the mixed Gaussian states.
Our aim is to fill this gap. 
The definition we propose for the mixed discrete Gaussian states is based on the explicit formulas we have obtained in the case of pure discrete variable Gaussian states.
\end{abstract}

\maketitle

The starting point in the description of quantum systems with finite-dimensional Hilbert space (also called finite quantum systems, discrete variable quantum systems or qudits) is the book of Weyl\cite{Weyl50} and the
remark of Schwinger \cite{Schwinger60} that two observables whose eigenstates are 
connected through the finite Fourier transform share a maximum degree of incompatibility.
A phase-space approach was initiated by Wootters \cite{Wootters87,Wootters89} and continued by Cohendet\cite{Cohendet88}, Galetti \cite{Galetti88}, Vourdas\cite{Vourdas90,Vourdas91,Vourdas04}, Hadzitaskos\cite{Hadzitaskos93}, Bu\v{z}ek\cite{Buzek95}, Opatrny \cite{Opatrny95,Opatrny96}, Leonhardt \cite{Leonhardt95,Leonhardt96}, Galetti\cite{Galetti96}, Tolar\cite{Tolar97}, Hakio\v{g}lu\cite{Hakioglu98}, Zhang\cite{Zhang04}, Ruzzi\cite{Ruzzi05},  Klimov \cite{Klimov09}, Marchiolli \cite{Marchiolli11,Marchiolli12,Marchiolli13}, DeBrota \cite{DeBrota20}, Rundle\cite{Rundle21} and many others.

A remarkable discrete version of Hermite-Gauss functions was presented by Mehta \cite{Mehta87} and
investigated in more details by Ruzzi \cite{Ruzzi06}.
The same method can be used in order to define Gaussian functions of discrete variable.
From a mathematical point of view, they are particular cases of the Jacobi function $\theta_3$.
Ruzzi\cite{Ruzzi06}, Galetti\cite{Ruzzi05}, Marcholli\cite{Marchiolli11}, Zhang and Vourdas\cite{Zhang04} have investigated these discrete Gaussian functions by using the 
properties of Jacobi $\theta$ functions. 
Some additional results have been obtained by Cotfas and Dragoman \cite{Cotfas12} by using an approach directly based on the computation with series.

The mathematical formalism available for the description
of quantum systems with finite-dimensional Hilbert space contains: 
finite discrete Fourier transform \cite{Mehta87,Ruzzi06,Vourdas04},
finite discrete position and momentum like operators \cite{Opatrny95,Opatrny96},
discrete position space \cite{Opatrny95,Opatrny96,Vourdas04},
discrete momentum space \cite{Opatrny95,Opatrny96,Vourdas04},
discrete phase-space \cite{Wootters87,Galetti96,Marchiolli11,Vourdas04} ,
discrete displacement operators \cite{Opatrny95,Opatrny96,Marchiolli11,Vourdas04},
discrete displaced parity (phase-point) operators \cite{Opatrny95,Opatrny96,Wootters87,Vourdas04,Rundle21},
discrete Wigner function \cite{Wootters87,Opatrny95,Opatrny96,DeBrota20},
discrete Weyl (characteristic) function \cite{Leonhardt95,Leonhardt96,Vourdas04},
discrete parameterized quasiprobability distributions \cite{Opatrny95,Opatrny96,Marchiolli11,Ruzzi05,DeBrota20},
discrete Husimi function \cite{Opatrny96,Ruzzi05},
discrete Glauber-Sudarshan function \cite{Opatrny96,Ruzzi05},
discrete quantum state tomography \cite{Leonhardt95,Leonhardt96},
discrete Weyl-Wigner-Moyal formalism \cite{Marchiolli12}
discrete Hermite-Gauss functions \cite{Mehta87,Galetti96,Ruzzi06},
Gaussian functions of discrete variable \cite{Ruzzi06},
discrete vacuum \cite{Marchiolli11,Ruzzi05}, 
discrete coherent states \cite{Galetti96,Marchiolli11,Rundle21,Klimov09,Ruzzi05,Cotfas10},
discrete squeezed states \cite{Klimov09},
discrete creation and annihilation operators \cite{Galetti96},
finite harmonic oscillator \cite{Galetti96,Hakioglu00,Lorente01,Atakishiev08,Cotfas11,Cotfas13},
discrete fractional Fourier transform \cite{Barker00,Ozaktas01,Wolf07,Cotfas13},
discrete Fock space \cite{Galetti96},
entropic uncertainty relation \cite{Vourdas04},
Harper functions \cite{Marchiolli13},
Fourier–Kravchuk transform \cite{Hakioglu00},
orthogonal polynomials of discrete variable\cite{Nikiforov91},
etc.

To our knowledge, a finite version of the mixed Gaussian states is missing, and our wish is to fill this gap. 
In the case of pure discrete variable Gaussian states we have obtained\cite{Cotfas12,Cotfas20} for the discrete Wigner function the formulas (\ref{Wig1old}) and (\ref{Wig2old}) (see Supplemental Material\cite{Supp}). The definition we propose for the mixed discrete variable Gaussian states is based on the remark (presented here for the first time) that these formulas can be written in the compact form (\ref{Wig1new}) and respectively (\ref{Wig2new}).
Then, we investigate some of the properties of the discrete Gaussian states defined in this way.
Particularly, we define for the first time (to our knowledge) discrete thermal states and a subspace of the Hilbert space where a discrete version of the position-momentum commutation relation is approximately satisfied.
The orthogonal projection of a discrete Gaussian state on this subspace seems to be much more significant then the projection on the orthogonal complement.

%
{\bf Phase-space formulation of finite quantum systems.-} 
In this section, we review the elements of the phase-space 
formulation of finite quantum systems we use in the next sections.
For simplicity,  we consider only quantum systems with 
an odd-dimensional ($d\!=\!2s\!+\!1$) Hilbert space $\mathcal{H}$, which can be regarded as the space of all the functions 
$\psi\!:\!\{-\!s,\!-s\!+\!1,...,s\!-\!1,s\}\!\rightarrow \!\mathbb{C}$
with the inner product
$
\langle \varphi|\psi\rangle \!=\!\sum\limits_{n=-s}^s\overline{\varphi(n)}\,\psi(n).
$
Each function $\psi\!\in\!\mathcal{H}$ can be extended up to a periodic function $\psi\!:\!\mathbb{Z}\!\rightarrow \!\mathbb{C}$
of period $d$, and $\mathcal{H}$ can also be regarded as the space of all the functions $ \psi\!:\!\mathbb{Z}_d\!\rightarrow \!\mathbb{C}$, that is $\mathcal{H}\!=\!\ell^2(\mathbb{Z}_d)$.
The canonical basis $\{ \delta_{-s},\ \delta_{-s+1},...,\delta_{s-1},\delta_s\}$, where
\begin{equation}
\delta_m(n)\!=\!\delta^{[d]}_{mn}\!=\!\left\{ \begin{array}{lll}
1  & \mbox{if} & n\!=\!m\ \ \mbox{modulo } d\\
0  & \mbox{if} & n\!\neq\!m\ \ \mbox{modulo } d
\end{array} \right.
\end{equation}
is an orthonormal basis. By using Dirac's notation $|m\rangle$ instead of $\delta _m$, we have
$
\langle m|k \rangle \!=\!\delta^{[d]}_{mk}$ and $ \sum\limits_{m=-s}^s|m\rangle\langle m|\!=\!\mathbb{I},
$
where $\mathbb{I}\!:\!\mathcal{H}\!\rightarrow\!\mathcal{H}, \ \, \mathbb{I}\psi\!=\!\psi,$ \, is the identity operator.\\ The discrete {\em Fourier transform} \cite{Mehta87,Vourdas04,Ruzzi06}
$\mathfrak{F}\!:\!\mathcal{H}\!\rightarrow\!\mathcal{H}\!:\!\psi \mapsto \mathfrak{F}[\psi]$,
\begin{equation}
\mathfrak{F}[\psi](k)\!=\!\mbox{\small $\frac{1}{\sqrt{d}}$}\!\sum\limits_{n=-s}^s {\rm e}^{-\frac{2\pi {\rm i}}{d}kn}\, \psi(n)
\end{equation}
is a unitary one: \ $\mathfrak{F}^{-1}\!=\!\mathfrak{F}^\dag$, \ $\mathfrak{F}^\dag[\psi](k)\!=\!\mbox{\small $\frac{1}{\sqrt{d}}$}\!\sum\limits_{n=-s}^s {\rm e}^{\frac{2\pi {\rm i}}{d}kn}\, \psi(n)$.\\
The self-adjoint operator ($-s\!\leq \mbox{\r{\it n}} \leq \!s$ is the representative modulo $d$ of $n$)
$
\hat{\mathfrak{q}}:\mathcal{H}\!\rightarrow\!\mathcal{H}\!:\!\psi \mapsto \hat{\mathfrak{q}}\psi$, \ 
$(\hat{\mathfrak{q}}\psi)(n)\!=\!\mbox{\r{\it n}}\, \psi(n)
$
corresponds to the {\em position operator}\cite{Galetti96,Vourdas04} $\hat{q}$, and
$
\hat{\mathfrak{p}}\!:\!\mathcal{H}\!\rightarrow\!\mathcal{H}$, \ $\hat{\mathfrak{p}}\!=\!\mathfrak{F}^\dag \hat{\mathfrak{q}}\mathfrak{F}
$
corresponds to the {\em momentum operator}\cite{Galetti96,Vourdas04} $\hat{p}$ from the continuous case. 
The operators (the addition is modulo $d$)
\begin{equation}
\begin{array}{ll} 
\mathcal{H}\!\rightarrow\!\mathcal{H}\!:\!\psi\!\mapsto\! \mathrm{ e}^{\frac{2\pi \mathrm{i}}{d}k \hat{\mathfrak{q}}}\psi,\quad  & \mathrm{e}^{\frac{2\pi {\rm i}}{d}k \hat{\mathfrak{q}}}\psi(n)\!=\!\mathrm{e}^{\frac{2\pi \mathrm{i}}{d}k n}\psi(n),\\[1mm]
\mathcal{H}\!\rightarrow\!\mathcal{H}\!:\!\psi\!\mapsto\!  {\rm e}^{-\frac{2\pi {\rm i}}{d}n \hat{\mathfrak{p}}}\psi, & {\rm e}^{-\frac{2\pi {\rm i}}{d}n \hat{\mathfrak{p}}}\psi(m)\!=\!\psi(m\!-\!n ),
\end{array}
\end{equation}
are the {\em translation operators}\cite{Schwinger60,Galetti96,Vourdas04}, and 
\begin{equation}
\mathfrak{D}(n ,k )\!:\!\mathcal{H}\!\rightarrow\!\mathcal{H},\quad 
\begin{array}{l}
\mathfrak{D}(n ,k )\!=\!{\rm e}^{-\frac{\pi {\rm i}}{d} n k }\,{\rm e}^{\frac{2\pi {\rm i}}{d}k \hat{\mathfrak{q}}}\, {\rm e}^{-\frac{2\pi {\rm i}}{d}n \hat{\mathfrak{p}}},\\[1mm]
\qquad \quad \  =\!\mathrm{e}^{\frac{\pi \mathrm{i}}{d} n k }\, \mathrm{e}^{-\frac{2\pi \mathrm{i}}{d}n \hat{\mathfrak{p}}}\,{\rm e}^{\frac{2\pi \mathrm{i}}{d}k \hat{\mathfrak{q}}},\\[1mm]
 \mathfrak{D}(n ,k )\psi (m)={\rm e}^{-\frac{\pi {\rm i}}{d} n k }\,{\rm e}^{\frac{2\pi {\rm i}}{d}km}\, \psi (m\!-\!n )
\end{array}
\end{equation}
are the {\em displacement operators}\cite{Opatrny95,Galetti96,Vourdas04}. The {\em displaced parity operators}\cite{Opatrny95,Galetti96,Vourdas04}
\begin{equation}
\Pi(n,k)\!:\!\mathcal{H}\!\rightarrow\!\mathcal{H},\quad 
\begin{array}{l}
\Pi(n ,k )\!=\!\mathfrak{D}(n ,k )\, \Pi \,\mathfrak{D}^\dag(n ,k )\\[1mm]
\Pi (n ,k )\psi (m)= {\rm e}^{-\frac{2\pi {\rm i}}{d}2k(n-m)}\, \psi (2n\!-\!m)
\end{array}
\end{equation}
where \ $\Pi \psi (n)=\psi(-n)$, form an orthogonal basis in the real Hilbert space 
$
\mathcal{A}(\mathcal{H})\!=\!\{\, A\!:\!\mathcal{H}\!\rightarrow\!\mathcal{H}\ |\ \ A^\dag\!=\!A\, \}
$
of all the self-adjoint operators, namely
\begin{equation}
\langle \Pi (n,k), \Pi(m,\ell)\rangle\!=\!{\rm tr}(\Pi (n,k)\, \Pi(m,\ell))\!=\!d\, \delta_{nm}\, \delta_{k\ell}.
\end{equation}
Any density operator \ $\bm{\varrho} \!:\!\mathcal{H}\!\rightarrow\!\mathcal{H}$ \ can be written as \cite{Leonhardt95}
\begin{equation}
\bm{\varrho}\!=\!\sum\limits_{n,k}\mathfrak{W}_{\bm{\varrho}}(n,k)\,\Pi (n,k),
\end{equation}
where (the addition is modulo $d$)
\begin{equation}
\begin{array}{l}
\mathfrak{W}_{\bm{\varrho}}\!:\!\{-\!s,\!-s\!+\!1,...,s\!-\!1,s\}\!\times\!\{-\!s,\!-s\!+\!1,...,s\!-\!1,s\}\!\longrightarrow\!\mathbb{R},\\[2mm]
\mathfrak{W}_{\bm{\varrho}}(n,k)\!=\!\frac{1}{d}\ {\rm tr}(\varrho\,\Pi (n,k))\!=\!\frac{1}{d}\,\sum\limits_{m=-s}^s {\rm e}^{-\frac{4\pi {\rm i}}{d}km}\, \langle n\!+\!m|\bm{\varrho}|n\!-\!m\rangle
\end{array}
\end{equation}
is the {\em discrete  Wigner function}\cite{Opatrny95,Galetti96,Vourdas04} of $\bm{\varrho}$. The Wigner function of a pure state $\bm{\varrho}\!=\!|\psi\rangle\langle \psi|$ is
\begin{equation}
\mathfrak{W}_{\psi}(n,\!k)\!=\!\frac{1}{\mbox{\footnotesize $d$}}\langle \psi|\Pi(n,k)|\psi\rangle \!=\!\frac{1}{\mbox{\footnotesize $d$}}\!\sum\limits_{m=-s}^s \!{\rm e}^{-\frac{4\pi {\rm i}}{d}km}\, \psi(n\!+\!m)\, \overline{\psi(n\!-\!m)}
\end{equation}
and has the marginal properties:
\[
\sum\limits_{k=-s}^s\mathfrak{W}_{\psi}(n,\!k)\!=\!|\psi(n)|^2,\qquad \sum\limits_{n=-s}^s\mathfrak{W}_{\psi}(n,\!k)\!=\!|\mathfrak{F}[\psi](k)|^2.
\]

%
{\bf Pure single-mode discrete variable Gaussian states.-}
\begin{figure}[t]
\includegraphics[scale=0.72]{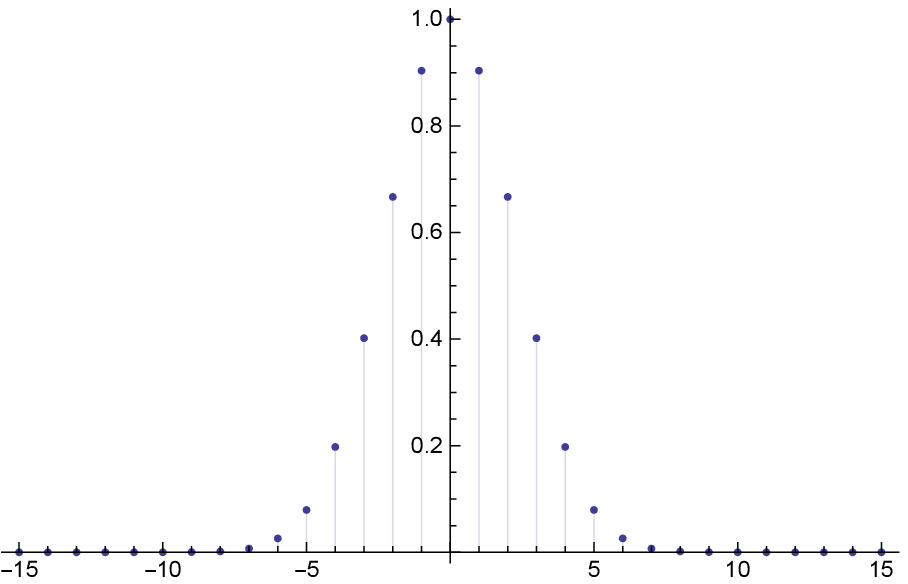}\\ 
\includegraphics[scale=0.55]{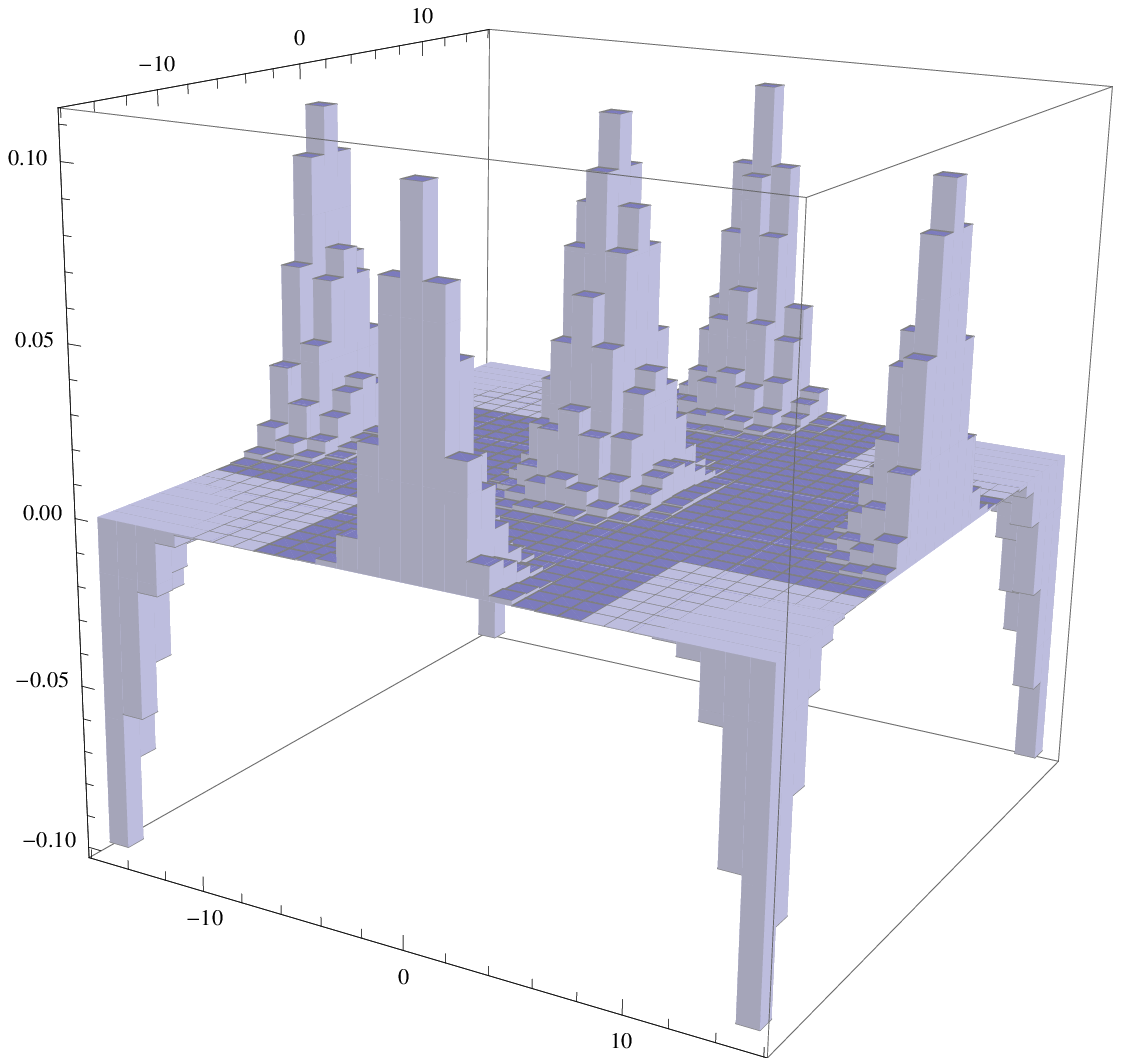}
\caption{The Gaussian function $\mathfrak{g}_1$ and its Wigner function for $d\!=\!31$. }
\end{figure}
The function  
$
g_\kappa :\mathbb{R}\rightarrow \mathbb{R}, \quad \mbox{\large $g_\kappa (q)\!=\!\mathrm{e}^{-\frac{\kappa}{2\hbar}q^2}\!=\!\mathrm{e}^{-\frac{\kappa\pi }{h}q^2}$}
$,where $\kappa \!\in\!(0,\infty)$ is a parameter,   
is a  {\em Gaussian function of continuous variable}. 
The Wigner function of \ $g_\kappa $ \ is \ $\mathcal{W}_{g_\kappa}\!:\!\mathbb{R}^2\!\rightarrow\!\mathbb{R}$,
\begin{equation}
\begin{array}{lll}
\mathcal{W}_{g_\kappa}(q,p)\!& \!\!\!\!= & \!\!\!\!\frac{2}{h}\int\limits_{-\infty}^{\infty} {\rm e}^{-\frac{4\pi{\rm i}}{h}px}\,g_\kappa(q\!+\!x)\, \overline{g_\kappa(q\!-\!x)}\, dx,\\[3mm]
\!& \!\!\!\!= & \!\!\!\!\sqrt{\frac{2}{\kappa h}}\ \mbox{\rm exp}\left\{-\frac{2\pi }{h}\mbox{\small  (\,$q$\ \ $p$\, )} \left(\!\!\!
\begin{array}{cc}
\mbox{\footnotesize $\kappa^{-1}$} & \mbox{\footnotesize 0}\\
\mbox{\footnotesize 0} & \mbox{\footnotesize $\kappa $}
\end{array}\!\!\!\right)^{\!\!-1} \!\!\!\left(\!\!\!
\begin{array}{c}
\mbox{\footnotesize $q$}\\
\mbox{\footnotesize $p$}
\end{array}\!\!\!\!\right)\right\}.
\end{array}
\end{equation}
The function\cite{Ruzzi06} (see Fig. 1)\ \ $ \mathfrak{g}_\kappa \!:\!\{-s,\!-s\!+\!\!1,...,s\!-\!\!1,\!s\}\!\rightarrow \!\mathbb{R},$
\begin{equation}
 \mathfrak{g}_\kappa(n)=\sum\limits_{\alpha=-\infty}^\infty{\rm e}^{-\frac{\kappa \pi}{d}(n+\alpha d)^2}
 =\sum\limits_{\alpha =-\infty }^\infty \!\!\!g_\kappa \!\left( (n\!+\!\alpha d)\mbox{\footnotesize{$\sqrt{\frac{h}{d}}$}} \right)
\end{equation}
can be regarded as a  {\em Gaussian function of discrete variable}.\\
The expressin of discrete Wigner function (see Fig. 1) 
\begin{equation}
\begin{array}{l}
\mathfrak{W}_{\mathfrak{g}_\kappa}\!:\!\{-\!s,\!-s\!+\!1,...,s\!-\!1,s\}\!\times\!\{-\!s,\!-s\!+\!1,...,s\!-\!1,s\}\!\longrightarrow\!\mathbb{R},\\[2mm]
\mathfrak{W}_{\mathfrak{g}_\kappa}(n,k)\!=\!\frac{1}{d}\!\sum\limits_{m=-s}^s\!{\rm e}^{-\frac{4\pi {\rm i}}{d}km}\, \mathfrak{g}_\kappa(n\!+\!m)\, \overline{\mathfrak{g}_\kappa(n\!-\!m)}.
\end{array}
\end{equation}
of \ $\mathfrak{g}_\kappa $ obtained by Cotfas and Dragoman \cite{Cotfas12} (formula  (\ref{Wig1old}) from Supplemental Material\cite{Supp}) can be written in the compact form
\begin{equation} \label{Wig1new}
\begin{array}{l}
\mathfrak{W}_{\!\mathfrak{g}_\kappa}(n,k)\!
=\!C\!\sum\limits_{\alpha,\beta  =-\infty }^\infty \!(-1)^{\alpha \beta}\ w\!\left( n\!+\!\mbox{\small{$\alpha \frac{d}{2}$}}, k\!+\!\mbox{\small{$\beta \frac{d}{2}$}}\right),
\end{array}
\end{equation}
where $C$ is a normalizing factor and
\begin{equation}\label{gkapp1}
w(q,p)\!=\! \ \mbox{\rm exp}\left\{-\frac{2\pi}{d}\mbox{\small  (\,$q$\ \ $p$)} \left(\!\!\!
\begin{array}{cc}
\mbox{\footnotesize $\kappa^{-1}$} & \!\mbox{\footnotesize 0}\\[-1mm]
\mbox{\footnotesize 0} & \!\mbox{\footnotesize $\kappa $}
\end{array}\!\!\!\right)^{\!\!-1} \!\!\!\left(\!\!\!
\begin{array}{c}
\mbox{\footnotesize $q$}\\
\mbox{\footnotesize $p$}
\end{array}\!\!\!\!\right)\right\}.
\end{equation}

%
%
%
%
{\bf Pure two-mode discrete variable Gaussian states.-}
Let $\tau\!=\! {\scriptsize \left(\!\!\begin{array}{cc}
a & \!b\\
b & \!c\end{array}\!\!\right)}$ be a $2\!\times\!2$ real symmetric positive-definite matrix. The real function $g_\tau\!:\!\mathbb{R}\!\times\!\mathbb{R}\!\longrightarrow \!\mathbb{R},$
\[
g_\tau(q_1,q_2)\!=\!\mathrm{exp}\left\{-\frac{\pi}{h}\mbox{\scriptsize  (\,$q_1$\ \ $q_2$)}\, {\scriptsize \left(\!\!\begin{array}{cc}
a & \!b\\
b & \!c\end{array}\!\!\right)}\,  {\scriptsize \left(\!\!\begin{array}{c}
q_1\\
q_2\end{array}\!\!\right)}\right\}
\]
is a  {\em Gaussian function of two continuous variables}. 
The Wigner function of \ $g_\tau $ \ is \ $\mathcal{W}_{g_\tau}\!:\!\mathbb{R}^4\!\rightarrow\!\mathbb{R}$, \ $\mathcal{W}_{g_\tau }(q_1,q_2,p_1,p_2)$
\[
=\frac{2}{h\sqrt{{\rm det}\ \tau}}\, {\rm exp}\left\{-\frac{2\pi }{h} \mbox{\scriptsize  (\,$q_1$\ \ $q_2$\ \ $p_1$\ \ $p_2$\, )}{\scriptsize \left(\!\!\!
\begin{array}{cc}
\tau^{-1} & \!\!\!\!\!\!\!\!\begin{array}{cc}
0 & \!0\\
0 & \!0\end{array}\\
\begin{array}{cc}
0 & \!0\\
0 & \!0\end{array} & \!\!\!\!\!\tau
\end{array}\!\!\!\right)^{\!\!\!-1} \!{\scriptsize \left(\!\!\!
\begin{array}{c}
q_1\\
q_2\\
p_1\\
p_2
\end{array}\!\!\!\!\right)}}\right\}.
\]
The function \\ $ \mathfrak{g}_\tau \!:\!\{-\!s,\!-s\!+\!1,...,s\}\!\times\!\{-\!s,\!-s\!+\!1,...,s\}\!\longrightarrow\!\mathbb{R},\ \ \mathfrak{g}_\tau(n_1,n_2)$
\begin{equation}
\begin{array}{l}
=\sum\limits_{\alpha_1,\alpha_2 =-\infty }^\infty 
\mathrm{exp}\left\{-\frac{\pi}{d}\mbox{\scriptsize  (\,$n_1\!+\!\alpha_1 d$\ \ \ $n_2\!+\!\alpha_2 d$)}\, {\scriptsize \left(\!\!\begin{array}{cc}
a & \!b\\
b & \!c\end{array}\!\!\right)}\,  {\scriptsize \left(\!\!\begin{array}{c}
n_1\!+\!\alpha_1d\\
n_2\!+\!\alpha_2 d\end{array}\!\!\right)}\right\}
\end{array}
\end{equation}
can be regarded as a  {\em Gaussian function of two discrete variables}. 
The explicit formula of the discrete Wigner function of \ $\mathfrak{g}_\tau $ \ obtained by Cotfas \cite{Cotfas20} (formula  (\ref{Wig2old}) from Supplemental Material\cite{Supp}) can be written in the compact form
\begin{equation}\label{Wig2new}
\begin{array}{l}
{\small \mathfrak{W}}_{\!\mathfrak{g}_{\tau}}(n_1,n_2,k_1,k_2)\!=\!C \sum\limits_{\alpha_1,\alpha_2 =-\infty }^\infty \ 
\sum\limits_{\beta_1,\beta_2 =-\infty }^\infty (-1)^{\alpha_1\beta_1+\alpha_2\beta_2}\\[4mm]
\qquad \quad \qquad \ \ \times w\!\left(n_1\!+\!\alpha_1\frac{d}{2},n_2\!+\!\alpha_2\frac{d}{2},k_1\!+\!\beta_1\frac{d}{2},k_2\!+\!\beta_2\frac{d}{2}\right),
\end{array}
\end{equation}
where $C$ is a normalizing factor, and
\begin{equation}\label{mixt2}
w(q_1,q_2,p_1,p_2)\!=\!\mathrm{exp}\left\{- \mbox{\small $\frac{2\pi }{d}$}\mbox{\scriptsize  (\,$q_1$\ \ $q_2$\ \ $p_1$\ \ $p_2$\, )}{\scriptsize \left(\!\!\!
\begin{array}{cc}
\tau^{-1} & \!\!\!\!\!\!\!\!\begin{array}{cc}
0 & \!0\\
0 & \!0\end{array}\\
\begin{array}{cc}
0 & \!0\\
0 & \!0\end{array} & \!\!\!\!\!\tau
\end{array}\!\!\!\right)^{\!\!\!-1} \!{\scriptsize \left(\!\!\!
\begin{array}{c}
q_1\\
q_2\\
p_1\\
p_2
\end{array}\!\!\!\!\right)}}\right\}.
\end{equation}

%
%
{\bf Pure and mixed discrete variable Gaussian states.-}
In the case of a continuous variable quantum system, 
a Gaussian state\cite{Weedbrook12,Ferraro05,Wang07,Adesso14,Navarrete22} $\rho$ is a state with a Wigner function of the form
\begin{equation}\label{wigner1}
\begin{array}{l}
\mathcal{W}_{\!\rho} (q,p)\!=\!\frac{2}{h \sqrt{\det\sigma}}\, {\rm exp}\left\{-\frac{2\pi }{h} (\,q-\tilde q\ \ p-\tilde p\, ){\scriptsize \left(\!\!\!
\begin{array}{cc}
\sigma_{11} & \sigma_{12}\\
\sigma_{12} & \sigma_{22}
\end{array}\!\!\!\right)^{\!\!-1} \!\!\!\left(\!\!\!
\begin{array}{c}
q\!-\!\tilde q\\
p\!-\!\tilde p
\end{array}\!\!\!\!\right)}\right\},
\end{array}
\end{equation}
where $\tilde q,\, \tilde p\!\in\!\mathbb{R}$ are two parameters, and the covariance matrix
\begin{equation}\label{sigcon}
\sigma\!=\!{\scriptsize \left(\!\!
\begin{array}{cc}
\sigma_{11} & \sigma_{12}\\
\sigma_{12} & \sigma_{22}
\end{array}\!\!\right)},\quad \mbox{satisfying}\quad \sigma \!+\!{\rm i} {\scriptsize \left(\!\!
\begin{array}{cc}
0 & 1\\
-1 & 0
\end{array}\!\!\right)}\!\geq \!0,
\end{equation}
is a real $2\!\times\!2$ symmetric positive-definite matrix. 
In the case $\tilde q\!=\!\tilde p\!=\!0$, we shall write $\rho_\sigma $ instead of $\rho$. 
Inspired by the relations (\ref{Wig1new}) and (\ref{gkapp1}), we propose the following definition.\\
{\it Definition 1:}\  In the case of a discrete variable quantum system, a state $\bm{\varrho}$ is a
{\it single-mode discrete variable Gaussian state} if there exists a real $2\!\times\!2$ symmetric positive-definite matrix 
$ \sigma\!=\!{\scriptsize \left(\!\!
\begin{array}{cc}
\sigma_{11} & \sigma_{12}\\
\sigma_{12} & \sigma_{22}
\end{array}\!\!\right)}
$ such that the Wigner function of $\bm{\varrho}$ is
\begin{equation}
\mathfrak{W}_{\!\bm{\varrho}}(n,k)\!
=\!C\!\sum\limits_{\alpha,\beta  =-\infty }^\infty \!(-1)^{\alpha \beta}\ w_\sigma\!\left( n\!+\!\mbox{\small{$\alpha \frac{d}{2}$}}, k\!+\! \mbox{\small{$\beta\frac{d}{2}$}}\right),
\end{equation}
where
\begin{equation}\label{wigner1}
\begin{array}{l}
w_\sigma(q,p)\!=\!{\rm exp}\left\{-\frac{2\pi}{d} (\,q-\tilde q\ \ p-\tilde p\, ){\scriptsize \left(\!\!\!
\begin{array}{cc}
\sigma_{11} & \sigma_{12}\\
\sigma_{12} & \sigma_{22}
\end{array}\!\!\!\right)^{\!\!-1} \!\!\!\left(\!\!\!
\begin{array}{c}
q\!-\!\tilde q\\
p\!-\!\tilde p
\end{array}\!\!\!\!\right)}\right\},
\end{array}
\end{equation}
$\tilde q,\, \tilde p\!\in\!\mathbb{R}$ are two parameters and $C$ is a normalizing constant.\\
For simplicity, such a state will also be called a {\it discrete Gaussian state} or a {\it finite Gaussian state}.\\
In the case $\tilde q\!=\!\tilde p\!=\!0$, we write $\bm{\varrho}_\sigma $ instead of $\bm{\varrho}$. We do not know what conditions $\sigma $ has to satisfy in order to have
\begin{equation}
\bm{\varrho}_{\sigma}\!=\!\sum\limits_{n,k}\mathfrak{W}_{\bm{\varrho}_\sigma}(n,k)\,\Pi (n,k)\!\geq\!0.
\end{equation}
Our numerical simulations suggest that (\ref{sigcon}) may be a sufficient condition.\\
{\it Theorem 1.}\  For any $\kappa \!\in\!(0,\infty)$, the pure state
$
{\bf g}_\kappa \!=\!\frac{1}{\sqrt{\langle \mathfrak{g}_\kappa,\mathfrak{g}_\kappa\rangle}}\, \mathfrak{g}_\kappa
$
 is\cite{Supp} a discrete variable Gaussian state with $\sigma\!=\!
\sigma_\kappa ={\scriptsize \left(\!\!
\begin{array}{cc}
\kappa^{-1} & \!\!0\\
0 & \!\!\kappa
\end{array}\!\!\right)}.
$
{\it Theorem 2.}\  For any $\kappa \!\in\!(0,\infty)$ and any $n_0,\, k_0\!\in\!\{-\!s,\!-s\!+\!1,...,s\!-\!1,s\}$,  the pure state
$
 \psi\!=\!\mathfrak{D}(n_0,k_0){\bf g}_\kappa 
$
is\cite{Supp} a discrete variable Gaussian state with $\sigma\!=\!\sigma_\kappa$.\\
{\it Theorem 3.}\ If $\bm{\varrho}$ is a discrete variable Gaussian state, then
 $\mathfrak{D}(n_0,k_0)\bm{\varrho}\mathfrak{D}^\dag(n_0,k_0)$ is\cite{Supp} also a discrete variable Gaussian state,
for any $n_0,\, k_0\!\in\!\{-\!s,\!-s\!+\!1,...,s\!-\!1,s\}$.\\
{\it Theorem 4.}\ The Fourier transform $\mathfrak{F}\bm{\varrho}\mathfrak{F}^\dag$ of a discrete variable Gaussian   state $\bm{\varrho}$  is\cite{Supp} also a discrete variable Gaussian state. If the covariance matrix corresponding to $\bm{\varrho}$  is $\sigma$, then the  matrix corresponding to $\mathfrak{F}\bm{\varrho}\mathfrak{F}^\dag$  is\cite{Supp} $\Omega \sigma \Omega^T$, where
{\small 
$
\Omega\!=\!\left(\begin{array}{rr}
0 & 1\\
-1 & 0
\end{array}\right).
$
}\\
The last two theorems show that  Fourier and displacement transforms preserve the nature of discrete variable Gaussian states, that is, they are {\em discrete Gaussian transforms}.

The  discrete Gaussian state
$
|0,0\rangle \!=\! |{\bf g}_1\rangle
$
corresponds to the {\em vacuum state}, and the discrete Gaussian states
$
|n,k\rangle \!=\!\mathfrak{D}(n,k)|{\bf g}_1\rangle
$
satisfying the resolution of the identity
$
\mbox{\small $\frac{1}{d}$}\sum\limits_{n,k=-s}^s\ |n,k\rangle \langle n,k|\!=\!\mathbb{I}
$,
correspond\cite{Galetti96,Cotfas10,Cotfas11} to the {\em canonical coherent states} from the continuous variable case.
Any pure state $\psi\!\in\!\mathcal{H}$ admits the representation
$
|\psi\rangle \!=\!\mathbb{I}|\psi\rangle\!=\!\mbox{\small $\frac{1}{d}$}\sum\limits_{n,k=-s}^s\ |n,k\rangle \langle n,k|\psi \rangle
$
 as a linear superposition of  discrete variable Gaussian states.\\
\mbox{\it Theorem 5.} \  Any continuous variable Gaussian state $\rho_\sigma$ is\cite{Supp} the limit of a sequence of discrete Gaussian states  $\bm{\varrho}_\sigma$:
\begin{equation}
 \qquad \quad \rho_\sigma\!=\!\lim\limits_{d\rightarrow \infty} \bm{\varrho}_\sigma.
\end{equation}

By following the suggestion offered by the formulas (\ref{Wig2new}) and (\ref{mixt2}), we propose the following definition.\\
{\it Definition 2:}\  In the case of a discrete variable quantum system, a state $\bm{\varrho}$ is a
{\it two-mode discrete variable Gaussian state} if there exists a real $4\!\times\!4$ symmetric positive-definite matrix 
$ \sigma $ such that the Wigner function of $\bm{\varrho}$ is
\begin{equation}
\begin{array}{l}
{\small \mathfrak{W}}_{\!\bm{\varrho}}(n_1,n_2,k_1,k_2)\!=\!C \sum\limits_{\alpha_1,\alpha_2 =-\infty }^\infty \ 
\sum\limits_{\beta_1,\beta_2 =-\infty }^\infty (-1)^{\alpha_1\beta_1+\alpha_2\beta_2}\\[4mm]
\qquad \quad \qquad \ \ \times w\!\left(n_1\!+\!\alpha_1\frac{d}{2},n_2\!+\!\alpha_2\frac{d}{2},k_1\!+\!\beta_1\frac{d}{2},k_2\!+\!\beta_2\frac{d}{2}\right),
\end{array}
\end{equation}
where $C$ is a normalizing constant and , up to a translation, 
\begin{equation}
w(q_1,q_2,p_1,p_2)\!=\!\mathrm{exp}\left\{- \mbox{\small $\frac{2\pi }{d}$}\mbox{\scriptsize  (\,$q_1$\ \ $q_2$\ \ $p_1$\ \ $p_2$\, )}\sigma^{-1} \!{\scriptsize \left(\!\!\!
\begin{array}{c}
q_1\\
q_2\\
p_1\\
p_2
\end{array}\!\!\!\!\right)}\right\}.
\end{equation}

%
%
%
{\bf Purity of a single-mode finite Gaussian state.-}
In the continuous case, the purity of $\rho_{\sigma}$ is
\[
{\rm tr}\, \rho_{\sigma}^2=2\pi \hbar\int_{\mathbb{R}^2} \mathcal{W}_{\rho_{\sigma}}^2(q,p)\, dqdp=\frac{1}{\sqrt{\det \sigma}}.
\]
In the discrete case, the {\it purity} is \ 
$
{\rm tr}\, \bm{\varrho}_\sigma^2\!=\!d\!\sum\limits_{n,k=-s}^s\!\! \mathfrak{W}_{\bm{\varrho}_\sigma}^2(n,k).
$\\
{\it Theorem 6}. For any covariance matrix $\sigma$, we have\cite{Supp}
\[
\begin{array}{l}
\lim\limits_{d\rightarrow \infty}{\rm tr}\, \bm{\varrho}_\sigma^2\!=\!\frac{1}{\sqrt{\det \sigma}}.
\end{array}
\]
\\
Generally, the convergence seems to be fast (see Table \ref{purity}).

 \begin{table}[t]
\caption{\label{purity}
Variation of the purity in four particular cases.
}
{\scriptsize
\begin{tabular}{cccc|cccc}
\hline
$\sigma$&  $\frac{1}{\sqrt{\det \sigma}} $ & $d$ &
${\rm tr}\, \bm{\varrho}_\sigma^2$ & $\sigma$&  $\frac{1}{\sqrt{\det \sigma}} $ & $d$ &
${\rm tr}\, \bm{\varrho}_\sigma^2$  \\
\colrule
$\left(\!\!
\begin{array}{cc}
2 & 0\\
0 & 2
\end{array}\!\!\right)$  & 0.5 &
$\begin{array}{c}
3\\
5\\
7\\
9
\end{array}$  & $\begin{array}{c}
0.52865\\
0.50099\\
0.50003\\
0.50000
\end{array}$     & 
$\left(\!\!
\begin{array}{cc}
1 & \!\!\sqrt{3}\\
\sqrt{3} & 6
\end{array}\!\!\right)$  & 0.5773 &
$\begin{array}{c}
3\\
5\\
7\\
9
\end{array}$  & $\begin{array}{c}
0.7020\\
0.5948\\
0.5795\\
0.5776
\end{array}$  \\
\colrule
$\left(\!\!
\begin{array}{cc}
3 & 2\\
2 & 2
\end{array}\!\!\right)$ &0.7071 &
$\begin{array}{c}
3\\
5\\
7\\
9
\end{array}$  &$\begin{array}{c}
0.6632\\
0.7009\\
0.7064\\
0.7070
\end{array}$      & 
$\left(\!\!
\begin{array}{cc}
7 & \!\!-\pi \\
-\pi & 5
\end{array}\!\!\right)$ & 0.19948 &
$\begin{array}{c}
3\\
5\\
7\\
9
\end{array}$  &
$\begin{array}{c}
0.3389\\
0.2332\\
0.2080\\
0.2017
\end{array}$ \\
\hline
\end{tabular}
}
\end{table}
\noindent{\it Theorem 7}. \ If the covariance matrix is such that ${\rm det}\, \sigma \!=\!1$, then 
$\bm{\varrho}_\sigma$ is\cite{Supp} a pure state, that is:\ \ ${\rm det}\, \sigma \!=\!1\ \Rightarrow \ {\rm tr}\, \bm{\varrho}_\sigma^2\!=\!1$.\\
In the analized cases (see Table \ref{purityvar}), one eigenvalue of $\bm{\varrho}_\sigma$ is 1 and all the others are almost null.

%
%
{\bf Single-mode discrete thermal states.-}
The Gaussian state of continuous variable with the covariance matrix
\begin{equation}\label{sigm}
\begin{array}{l}
I_\nu=\left(\!\!
\begin{array}{cc}
\nu  & 0\\
0 & \nu
\end{array}\!\!\right) \qquad \mbox{with}\ \ \nu\!>\!1,
\end{array}
\end{equation}
is the thermal state \cite{Weedbrook12,Ferraro05,Wang07,Adesso14,Navarrete22}
$
\begin{array}{l}
\rho_{I_\nu} =\frac{2}{\nu \!+\!1}\sum\limits_{n=0}^\infty \left(\frac{\nu \!-\!1}{\nu \!+\!1}\right)^n|\Psi_n\rangle\langle \Psi_n|,
\end{array}
$
where   
$
\Psi_n(q)\!=\!\frac{1}{\sqrt{n!2^n\sqrt{\pi}}}H_n(q){\rm e}^{-\frac{q^2}{2}}
$
is the Hermite-Gauss function. Particularly, this means that the spectrum of $\rho_{I_\nu}$ is formed by the numbers
\[
\begin{array}{l}
\frac{1}{\sum\limits_{k=0}^\infty \left(\frac{\nu -1}{\nu +1}\right)^k}\left(\frac{\nu \!-\!1}{\nu \!+\!1}\right)^n,\qquad \mbox{where}\ \ n\!\in\!\{0,1,2,3,...\}.
\end{array}
\]
The spectrum of the discrete Gaussian state $\bm{\varrho}_{I_\nu}$ with the covariance matrix $I_\nu$ is well-approximated by the set of numbers
\[
\begin{array}{l}
N_n\!=\!\frac{1}{\sum\limits_{k=0}^{d-1} \left(\frac{\nu -1}{\nu +1}\right)^k}\left(\frac{\nu \!-\!1}{\nu \!+\!1}\right)^n\!\!\!,\ \mbox{where} \ n\!\in\!\{0,1,2,3,...,d\!-\!1\}.
\end{array}
\]
For example, in the case $\nu \!=\!2$, $d\!=\!7$, the spectrum is\\
0.6667, \ 0.2219,\ 0.0751,\ 0.0229,\ 0.0105,\ 0.0017,\ 0.0010,\\
and the numbers $N_0$, $N_1$, ..., $N_6$ are\\
0.6669,\ 0.2223,\ 0.0741,\ 0.0247,\ 0.0082,\ 0.0027, 0.0009.\\
 \begin{table}[t]
\caption{\label{purityvar}
Some discrete variable Gaussian states $\bm{\varrho}_\sigma $ with ${\rm det}\sigma \!=\!1$.
} 
{\scriptsize
\begin{tabular}{cl}
\hline
 $\sigma$ & $d$\qquad \quad Spectrum of the density operator $\bm{\varrho}_\sigma $\\
\hline
$\left(\!\!
\begin{array}{cc}
2 & 1\\
1 & 1
\end{array}\!\!\right)$ &
 $\begin{array}{l}
3\qquad  \begin{array}{l}
1., \ -7.26023*10^{-17}, \ 4.48467*10^{-17}
\end{array}\\[2mm]
5\qquad  \begin{array}{l}
1.,\ \  1.38888*10^{-16},\  8.37904*10^{-17},\\
\quad   -5.87708*10^{-17}, \ -6.19389*10^{-18}
\end{array} 
\\[4mm]
7\qquad  \begin{array}{l}
1.,\ \ 1.58*10^{-16},  -7.84*10^{-17}, -6.73*10^{-17},\\ 
\quad \ \  5.24*10^{-17}, \ 3.19*10^{-17}, \ -3.10*10^{-17}
\end{array}
\end{array} $ \\
\hline
$\left(\!\!
\begin{array}{cc}
1 & 3\\
3 & 10
\end{array}\!\!\right)$ & $\begin{array}{l}
3\qquad  \begin{array}{l}
1., \ \ 4.47648*10^{-17},\ 1.68388*10^{-17}
\end{array}\\[2mm]
5\qquad  \begin{array}{l}
1.,\ -9.69875*10^{-17},\ 8.67892*10^{-17},\\
\quad \ \  -3.28385*10^{-17},\ 1.97939*10^{-17}
\end{array} 
\\[4mm]
7\qquad  \begin{array}{l}
1.,\ \  3.21*10^{-16},\ 9.10*10^{-17}, -8.99*10^{-17},\\
\quad \ \ 4.99*10^{-17}, -3.57*10^{-17},\ 2.98*10^{-17}
\end{array}
\end{array} $ \\
\hline
$\left(\!\!
\begin{array}{cc}
3 & \sqrt{5}\\
\sqrt{5} & 2
\end{array}\!\!\right)$  &
 $\begin{array}{l}
3\qquad  \begin{array}{l}
1., \ \ 2.17738*10^{-16}, \ 3.20619*10^{-17}
\end{array}\\[2mm]
5\qquad  \begin{array}{l}
1., \ \ 2.70799*10^{-16},\ 1.12761*10^{-16},\\
\quad \ \ -9.79558*10^{-17}, \ -2.03061*10^{-17}
\end{array} 
\\[4mm]
7\qquad  \begin{array}{l}
1., \ \ 2.37*10^{-16}, -1.31*10^{-16}, \ 1.17*10^{-16},\\
\quad \ -8.50*10^{-17}, -2.36*10^{-17}, -1.01*10^{-18}
\end{array}
\end{array} $
\\ 
\hline
$\left(\!\!\!
\begin{array}{cc}
3 & \!\!-\sqrt{5}\\
-\sqrt{5} & \!\!2
\end{array}\!\!\right)$ &
 $\begin{array}{l}
3\qquad  \begin{array}{l}
1., \ 2.28273*10^{-16},\ 7.70378*10^{-17}
\end{array}\\[2mm]
5\qquad  \begin{array}{l}
1.,\ \  2.85697*10^{-16}, \ -8.77195*10^{-17},\\
\quad \ \ \  8.2357*10^{-17},\ -4.52847*10^{-17}
\end{array} 
\\[4mm]
7\qquad  \begin{array}{l}
1.,\ \  2.63*10^{-16}, -1.46*10^{-16},\ 1.30*10^{-16},\\ \quad \   -4.82*10^{-17}, -1.63*10^{-17}, -4.67*10^{-19}
\end{array}
\end{array} $\\
\hline
\end{tabular}
}
\end{table}
\noindent{\it Definition 3:} We call a {\em discrete thermal state} any discrete Gaussian state with a covariance matrix of the form (\ref{sigm}).

Since $\mathfrak{W}_{\bm{\varrho}_{I_\nu}}(n,-k)\!=\!\mathfrak{W}_{\bm{\varrho}_{I_\nu}}(n,k)\!=\!\mathfrak{W}_{\bm{\varrho}_{I_\nu}}(k,n)$, we have
\[
\begin{array}{r}
\overline{\langle \delta_a|\bm{\varrho}_{I_\nu}|\delta_b\rangle}\!=\!\langle \delta_a|\bm{\varrho}_{I_\nu}|\delta_b\rangle\!=\!\langle \delta_b|\bm{\varrho}_{I_\nu}|\delta_a\rangle.
\end{array}
\]
 Consequently, in $\ell^2(\mathbb{Z}_d)$ there exists an orhonormal basis $\{|n\rangle\!\rangle\} $ formed by eigenfunctions of
$\bm{\varrho}_{I_\nu}$ with real values. 

In the case $d\!=\!3$, we have\cite{Supp} $\bm{\varrho}_{I_\nu}\, \bm{\varrho}_{I_\mu}=\bm{\varrho}_{I_\mu}\, \bm{\varrho}_{I_\nu}$, for any $\nu\!>\!1$ and $\mu\!>\!1$. In the case $d\!>\!3$, our  numerical data show\cite{Supp} that 
$\bm{\varrho}_{I_\nu}\, \bm{\varrho}_{I_\mu}\!\approx\!\bm{\varrho}_{I_\nu}\, \bm{\varrho}_{I_\nu}$, and the eigenfunctions of all the density operators $\bm{\varrho}_{I_\nu}$ are almost the same.  Therefore, the real eigenfunctions $|n\rangle\!\rangle $,
 considered in the increasing order of the number of sign alternations,  can be regarded as a discrete version of the Hermite-Gauss functions, and
\[
\begin{array}{l}
H=\sum\limits_{n=0}^{d-1} \left(n\!+\!\frac{1}{2}\right) | n\rangle\!\rangle \langle\!\langle n|
\end{array}
\]
as the Hamiltonian of a $d$-dimensional counterpart of the harmonic quantum oscillator.

%
%
%
{\bf On the position-momentum commutation relation.-}
In the continuous case, the usual position-momentum commutation relation
\begin{equation}
[\hat{q},\hat{p}]={\rm i}\mbox{\small $\frac{h}{2\pi }$}
\end{equation}
is satisfied in a subspace dense in $L^2(\mathbb{R})$. The discrete counterpart
\begin{equation}
[\hat{\mathfrak{q}},\hat{\mathfrak{p}}]={\rm i}\mbox{\small $\frac{d}{2\pi }$}
\end{equation}
is not satisfied, but for $d$ large enough, most of the eigenvalues of the operator $[\hat{\mathfrak{q}},\hat{\mathfrak{p}}]-{\rm i}\frac{d}{2\pi }$ are almost null. For example,  in the cases $d\!=\!11$ and $d\!=\!61$, 
the eigenvalues of $[\hat{\mathfrak{q}},\hat{\mathfrak{p}}]-{\rm i}\frac{d}{2\pi }$ are:
{\scriptsize 
\begin{equation}
\begin{array}{ll}
 \underline{\mbox{Case }d=11} &\hspace{3cm} \underline{\mbox{Case }d=61} \\[1mm]
\!\!\!\!\!\begin{array}{l}
\ -1.4\!*\!10^{-8}{\rm i}\\
\ \ \ 7.9\!*\!10^{-7}{\rm i}\\
\ \ \ 2\!*\!10^{-5}{\rm i}\\
\ \ \ 3.3\!*\!10^{-4}{\rm i}\\
\ -3.9\!*\!10^{-3}{\rm i}\\
\ \ \ 3.4\!*\!10^{-2}{\rm i}\\
\ - 0.24\, {\rm i}\\
\ \ \ 1.33\, {\rm i}\\
\ - 5.45\, {\rm i}\\
\ \ 19.99\, {\rm i}\\
- 34.92\, {\rm i}\\
\end{array} & 
\!\!\!\!\!\left|\begin{array}{rrrr}
-4.0*10^{-15}{\rm i},\ & -3.4*10^{-14}{\rm i},\ & 8.7*10^{-14}{\rm i} ,\ & 0.00007\,{\rm i}\\
9.3*10^{-15}{\rm i},\ & -3.7*10^{-14}{\rm i} ,\ & -8.9*10^{-14}{\rm i}  ,\ &  -0.00030\,{\rm i} 
\\
7.4*10^{-15}{\rm i},\ &  4.1*10^{-14}{\rm i} ,\ &  -9.9*10^{-14}{\rm i},\ &   0.00127\, {\rm i}  
\\
-1.1*10^{-15}{\rm i},\ & -4.4*10^{-14}{\rm i} ,\ &  1.1*10^{-13}{\rm i},\ &  -0.00513\,{\rm i} 
\\
-3.7*10^{-15}{\rm i},\ & 4.0*10^{-14}{\rm i} ,\ &  -5.7*10^{-13}{\rm i},\ &  0.01981\, {\rm i}  
\\
1.1*10^{-14}{\rm i},\ & 4.7*10^{-14}{\rm i},\ &  3.6*10^{-12}{\rm i},\ &  -0.07335\,{\rm i}  \\
-1.6*10^{-14}{\rm i},\ & -4.9*10^{-14}{\rm i} ,\ &  -2.3*10^{-11}{\rm i},\ &  0.26040\, {\rm i} 
\\
-2.3*10^{-14}{\rm i},\ & -4.6*10^{-14}{\rm i} ,\ &  1.4*10^{-10}{\rm i},\ & -0.88457\,{\rm i}   
\\
2.2*10^{-14}{\rm i},\ & 6.8*10^{-14}{\rm i} ,\ &  -8.3*10^{-10}{\rm i},\ &  2.87316\, {\rm i}  
\\
2.7*10^{-14}{\rm i},\ & -6.8*10^{-14}{\rm i} ,\ &  4.7*10^{-9}{\rm i},\ & -8.86191\, {\rm i} \\
2.7*10^{-14}{\rm i},\ & 6.8*10^{-14}{\rm i} ,\ &  -2.5*10^{-8}{\rm i},\ & 26.1438\, {\rm i} 
\\
-3.3*10^{-14}{\rm i},\ & 7.0*10^{-14}{\rm i} ,\ &  1.3*10^{-7}{\rm i},\ & -70.9696\, {\rm i}  \\
-3.3*10^{-14}{\rm i},\ & -7.2*10^{-14}{\rm i} ,\ & -6.8*10^{-7}{\rm i},\ &  189.607\, {\rm i} \\
3.1*10^{-14}{\rm i},\ & 8.0*10^{-14}{\rm i} ,\ & 3.3*10^{-6}{\rm i},\ & -404.509\,{\rm i} \\
3.6*10^{-14}{\rm i},\ & -8.1*10^{-14}{\rm i} ,\ & -0.00001\, {\rm i},\ & 944.717\,{\rm i} \\
 & & &  -1270.53\, {\rm i}
 
\end{array}\right.
\end{array}
\end{equation}
}


Let $\lambda _1$, $\lambda _2$, ..., $\lambda _d$ be the  eigenvalues of $[\hat{\mathfrak{q}},\hat{\mathfrak{p}}]-{\rm i}\frac{d}{2\pi }$, considered in the increasing order of their modulus ($|\lambda _1|\leq |\lambda _2|\leq ...\leq |\lambda _d|$), and let
$\varphi_1$, $\varphi_2$, ... , $\varphi_d$  be the corresponding eigenfunctions.
If we expand a discrete variable Gaussian state in terms of the eigenbasis $\{ \varphi_k\}$, the most significant coefficients seem\cite{Supp} to be those corresponding to the functions $\varphi_k$ with small $|\lambda_k|$.
For $\varepsilon \!>\!0$, let $d_\varepsilon$ be such that 
$\{ \lambda _1,\, \lambda _2,\, ...\, ,\, \lambda _{d_\varepsilon}\}\!=\!\{\, \lambda_k\ |\ \ |\lambda_k|\!<\!\varepsilon\,\},$
and let \
$\mathcal{H}_\varepsilon\!=\!{\rm span}\{\,\varphi_1,\, \varphi_2,\, ...\, ,\, \varphi_{d_\varepsilon}\}.$\\
If $\varepsilon$ is small enough, then we can consider that 
\begin{equation}
[\hat{\mathfrak{q}},\hat{\mathfrak{p}}]\approx{\rm i}\mbox{\small $\frac{d}{2\pi }$}\quad \mbox{in}\quad \mathcal{H}_\varepsilon.
\end{equation}
By following the analogy with the continuous case, we define
\begin{equation}
\begin{array}{r}
\hat{\mathfrak{a}}\!=\!\sqrt{\frac{\pi}{d}}(\hat{\mathfrak{q}}\!+\!{\rm i}\hat{\mathfrak{p}}),\\[2mm]
\hat{\mathfrak{a}}^\dag\!=\!\sqrt{\frac{\pi}{d}}(\hat{\mathfrak{q}}\!-\!{\rm i}\hat{\mathfrak{p}}),
\end{array}\quad \mbox{similar to}\quad 
\begin{array}{r}
\hat{a}\!=\!\sqrt{\frac{\pi}{h}}(\hat{q}\!+\!{\rm i}\hat{p}),\\[2mm]
\hat{a}^\dag\!=\!\sqrt{\frac{\pi}{h}}(\hat{q}\!-\!{\rm i}\hat{p}),
\end{array}
\end{equation}
The discrete variable Gaussian states seem\cite{Supp} to mainly belong  to $\mathcal{H}_\varepsilon$ and $\mathcal{A}(\mathcal{H}_\varepsilon)$.
In the case of continuous variable, the transformations 
$\hat U\!=\! {\rm e}^{-\frac{\rm i}{2}\hat{H}}$, where
\begin{equation}
\begin{array}{l}
\hat{H}\!=\! (\,\hat{a}^\dag \ \ \hat{a}\,){\small \left(\!\!
\begin{array}{cc}
A & \!\!B\\
\bar{B} &\!\! A
\end{array}\!\!\right)}
\left(\!\!
\begin{array}{c}
\hat{a}\\
\hat{a}^\dag
\end{array}\!\!\right)\quad \mbox{with}\ \ 
\mbox{\small $\begin{array}{r}
A\!\in\!\mathbb{R},\\
B\!\in\!\mathbb{C},
\end{array}$}
\end{array}
\end{equation}
are Gaussian transforms, that is they preserve the Gaussian nature of the states.
For $\hat U\!=\!{\rm e}^{-\frac{\rm i}{2}\hat{H}}$,   there exists 
  $S\!\in\!{\rm Sp}(2,\mathbb{R})$ such that
\begin{equation}
{\rm e}^{-\frac{\rm i}{2}\hat{H}}\rho_\sigma {\rm e}^{\frac{\rm i}{2}\hat{H}} =\rho_{S\sigma S^T}.
\end{equation}

In the case of discrete variable, the relation $\lim\limits_{d\rightarrow \infty}\bm{\varrho}_\sigma=\rho_\sigma$ suggests that for
\begin{equation}
\begin{array}{l}
\hat{\mathfrak{H}}\!=\! (\,\hat{\mathfrak{a}}^\dag \ \ \hat{\mathfrak{a}}\,){\small \left(\!\!
\begin{array}{cc}
A & \!\!B\\
\bar{B} &\!\! A
\end{array}\!\!\right)}
\left(\!\!
\begin{array}{c}
\hat{\mathfrak{a}}\\
\hat{\mathfrak{a}}^\dag
\end{array}\!\!\right)\quad \mbox{with}\ \ 
\mbox{\small $\begin{array}{l}
A\!\in\!\mathbb{R},\\
B\!\in\!\mathbb{C},
\end{array}$}
\end{array}
\end{equation}
we must have
\begin{equation}\label{discGaussTr}
{\rm e}^{-\frac{\rm i}{2}\hat{\mathfrak{H}}}\bm{\varrho}_\sigma {\rm e}^{\frac{\rm i}{2}\hat{\mathfrak{H}}} \approx\bm{\varrho}_{S\sigma S^T}\qquad \mbox{for}\ \ d\ \ \mbox{large enough}.
\end{equation}
Some numerical results concerning a phase shift and a single-mode squeezing transformation are presented in Table III.

 \begin{table}[h]
\caption{\label{unitGauss}
Norm of  ${\rm e}^{-\frac{\rm i}{2}\hat{\mathfrak{H}}}\bm{\varrho}_\sigma {\rm e}^{\frac{\rm i}{2}\hat{\mathfrak{H}}}\!-\!\bm{\varrho}_{S\sigma S^T}$ in two particular cases. 
}
{\scriptsize
\qquad\qquad\qquad\qquad\begin{tabular}{ccccc}
\hline
$A$ &  $B$ & $\sigma$ & $d$ & $\left|\!\left| {\rm e}^{-\frac{\rm i}{2}\hat{\mathfrak{H}}}\bm{\varrho}_\sigma {\rm e}^{\frac{\rm i}{2}\hat{\mathfrak{H}}}\!-\!\bm{\varrho}_{S\sigma S^T}\right|\!\right|$\\
\hline
$\frac{\pi}{4}$ & $0$ & $\left(\!\!
\begin{array}{cc}
2 & 1\\
1 & 1
\end{array}\!\!\right)$  &
$\begin{array}{c}
5\\
7\\
9\\
11\\
13\\
15
\end{array}$  & $\begin{array}{c}
0.1701\\
0.0932\\
0.0489\\
0.0256\\
0.0135\\
0.0071
\end{array}$\\  
\hline
$0$ & $\frac{1}{2}\, {\rm e}^{{\rm i}\frac{\pi}{3}}$ & $\left(\!\!
\begin{array}{cc}
3 & 2\\
2 & 2
\end{array}\!\!\right)$  &
$\begin{array}{c}
5\\
7\\
9\\
11\\
13\\
15
\end{array}$  & $\begin{array}{c}
0.1605\\
0.1217\\
0.1057\\
0.0887\\
0.0729\\
0.0593
\end{array}$\\  
\hline
\end{tabular}
}
\end{table}
\noindent A possible explanation can be obtained\cite{Supp} by following the analogy with the continuous case\cite{Navarrete22}.

%
{\bf Conclusion and outlook.-}
The use of the discrete variable Gaussian states extends the class of the quantum states which can be described analytically. These particular states have some significant properties and may be useful in certain applications. They  can also be used as finite-dimensional approximates for the continuous variable Gaussian states.
Generally, in $ \rho_\sigma\!=\!\lim\limits_{d\rightarrow \infty} \bm{\varrho}_\sigma$, the convergence is very fast.
\newpage
\begin{widetext}
\begin{center}
 \textbf{\large Supplemental Material -- Mixed discrete variable Gaussian states} \\[2mm]
 \textbf{CONTENTS}
\end{center}
I. Pure single-mode discrete variable Gaussian states\hfill 6\\
II. Pure two-mode discrete variable Gaussian states\hfill 7\\
III. Pure and mixed discrete variable Gaussian states\hfill 8\\
IV. Position-momentum commutation relation and the discrete variable Gaussian states\hfill 10\\
\begin{center}
 \textbf{\large I. Pure single-mode discrete variable Gaussian states}
\end{center}
The  discrete Fourier transform  $ \mathfrak{F}[\mathfrak{g}_\kappa] \!:\!\{-s,\!-s\!+\!\!1,...,s\!-\!\!1,\!s\}\!\rightarrow \!\mathbb{R},$
\begin{equation}
\begin{array}{lll}
\mathfrak{F}[\mathfrak{g}_\kappa](k)\!& \!\!\!\!= & \!\!\!\!\frac{1}{\sqrt{d}}\!\sum\limits_{n=-s}^s\! {\rm e}^{-\frac{2\pi {\rm i}}{d}kn}\, \mathfrak{g}_\kappa(n)\\[4mm]
 \!& \!\!\!\!= & \!\!\!\!\frac{1}{\sqrt{\kappa}}\sum\limits_{\alpha=-\infty}^\infty{\rm e}^{-\frac{\pi}{\kappa d}(k+\alpha d)^2}\\[4mm]
 \!& \!\!\!\!= & \!\!\!\!\frac{1}{\sqrt{\kappa}}\sum\limits_{\alpha =-\infty }^\infty \!\!\!g_{\kappa^{\mbox{\tiny $-1$}}} \!\left( (k\!+\!\alpha d)\mbox{\footnotesize{$\sqrt{\frac{h}{d}}$}} \right)\\[4mm]
 \!& \!\!\!\!= & \!\!\!\!\sum\limits_{\alpha =-\infty }^\infty \!\!\!\mathcal{F}[g_\kappa] \!\left( (k\!+\!\alpha d)\mbox{\footnotesize{$\sqrt{\frac{h}{d}}$}} \right)
\end{array}
\end{equation}
of $\mathfrak{g}_\kappa$ is also a Gaussian function of discrete variable, and\cite{Cotfas12}
\begin{equation}
\mathfrak{F}[\mathfrak{g}_\kappa]\!=\!\mbox{\small $\frac{1}{\sqrt{\kappa}}$}\ \mathfrak{g}_{\kappa^{\mbox{\tiny $-1$}}}.
\end{equation}
The discrete Wigner function of \ $\mathfrak{g}_\kappa $ \ is 
\ $\mathfrak{W}_{\mathfrak{g}_\kappa}\!:\!\{-\!s,\!-s\!+\!1,...,s\!-\!1,s\}\!\times\!\{-\!s,\!-s\!+\!1,...,s\!-\!1,s\}\!\longrightarrow\!\mathbb{R},$
\begin{equation}\label{Wig1old}
\begin{array}{l}
\mathfrak{W}_{\mathfrak{g}_\kappa}(n,k)\!=\!\frac{1}{d}\!\sum\limits_{m=-s}^s\!{\rm e}^{-\frac{4\pi {\rm i}}{d}km}\, \mathfrak{g}_\kappa(n\!+\!m)\, \overline{\mathfrak{g}_\kappa(n\!-\!m)}\\
\qquad \qquad \ =\ \frac{1}{\sqrt{2\kappa d}}\mathfrak{g}_{2\kappa}(n)
\left(\mathfrak{g}_{\frac{2}{\kappa}}(k)+\mathfrak{g}_{\frac{2}{\kappa}}^+(k)\right)\\
\qquad \qquad \quad +\frac{1}{\sqrt{2\kappa d}}\mathfrak{g}_{2\kappa}^+(n)
\left(\mathfrak{g}_{\frac{2}{\kappa}}(k)-\mathfrak{g}_{\frac{2}{\kappa}}^+(k)\right),
\end{array}
\end{equation}
where
\ $ \mathfrak{g}_\kappa^+ \!:\!\{-s,\!-s\!+\!\!1,...,s\!-\!\!1,\!s\}\!\rightarrow \!\mathbb{R},$
\begin{equation}
\begin{array}{lll}
 \mathfrak{g}_\kappa^+(n)\!& \!\!\!\!= & \!\!\!\!\sum\limits_{\alpha=-\infty}^\infty{\rm e}^{-\frac{\kappa \pi}{d}(n+(\alpha+\frac{1}{2}) d)^2}\\
 \!& \!\!\!\!= & \!\!\!\!\sum\limits_{\alpha =-\infty }^\infty \!\!\!g_\kappa \!\left( (n\!+\!(\alpha+\frac{1}{2}) d)\mbox{\footnotesize{$\sqrt{\frac{h}{d}}$}} \right).
\end{array}
\end{equation}
This formula obtained  by Cotfas and Dragoman \cite{Cotfas12} can be written as
\begin{equation} \label{gkappa}
\begin{array}{l}
\mathfrak{W}_{\!\mathfrak{g}_\kappa}(n,k)\!
=\!C\!\sum\limits_{\alpha,\beta  =-\infty }^\infty \!(-1)^{\alpha \beta}\ w\!\left( n\!+\!\mbox{\small{$\alpha \frac{d}{2}$}}, k\!+\!\mbox{\small{$\beta \frac{d}{2}$}}\right),
\end{array}
\end{equation}
where $C$ is a normalizing factor and
\begin{equation}\label{gkappa1}
w(q,p)\!=\! \mbox{\large \rm e}^{-\frac{2\pi}{d}\mbox{\small  (\,$q$\ \ $p$)} \left(\!\!\!
\begin{array}{cc}
\mbox{\footnotesize $\kappa^{-1}$} & \!\mbox{\footnotesize 0}\\[-1mm]
\mbox{\footnotesize 0} & \!\mbox{\footnotesize $\kappa $}
\end{array}\!\!\!\right)^{\!\!-1} \!\!\!\left(\!\!\!
\begin{array}{c}
\mbox{\footnotesize $q$}\\
\mbox{\footnotesize $p$}
\end{array}\!\!\!\!\right)},
\end{equation}
that is
\begin{equation}
\begin{array}{l}
\mathfrak{W}_{\!\mathfrak{g}_\kappa}\!(n,k)\!
=\!C\sqrt{\frac{\kappa h}{2}}\!\!\sum\limits_{\alpha,\beta  =-\infty }^\infty \!\!\!(-1)^{\alpha \beta}\mathcal{W}_{g_\kappa}\!\!\left( \!(n\!+\!\mbox{\small{$\alpha \frac{d}{2}$}})\mbox{\footnotesize{$\sqrt{\frac{h}{d}}$}}, (k\!+\! \mbox{\small{$\beta \frac{d}{2}$}})\mbox{\footnotesize{$\sqrt{\frac{h}{d}}$}}\right).
\end{array}
\end{equation}
\mbox{}\\
\begin{center}
 \textbf{\large II. Pure two-mode discrete variable Gaussian states}
\end{center}
The function \ $ \mathfrak{g}_\tau \!:\!\{-\!s,\!-s\!+\!1,...,s\!-\!1,s\}\!\times\!\{-\!s,\!-s\!+\!1,...,s\!-\!1,s\}\!\longrightarrow\!\mathbb{R},$
\begin{equation}
\begin{array}{lll}
\mathfrak{g}_\tau(n_1,n_2)\!& \!\!\!\!= & \!\!\!\!\sum\limits_{\alpha_1,\alpha_2 =-\infty }^\infty 
\mathrm{exp}\left\{-\frac{\pi}{d}{\scriptsize \mbox{$(n_1\!+\!\alpha_1 d\ \ \ n_2\!+\!\alpha_2 d)$}}\, {\scriptsize \left(\!\!\begin{array}{cc}
a & \!b\\
b & \!c\end{array}\!\!\right)}\,  {\scriptsize \left(\!\!\begin{array}{c}
n_1\!+\!\alpha_1d\\
n_2\!+\!\alpha_2 d\end{array}\!\!\right)}\right\}\\
 \!& \!\!\!\!= & \!\!\!\!\sum\limits_{\alpha_1,\alpha_2 =-\infty }^\infty\!\!\! g_\tau\left((n_1\!+\!\alpha _1 d)\sqrt{\frac{h}{d}}, (n_2\!+\!\alpha _2 d)\sqrt{\frac{h}{d}}\right)
\end{array}
\end{equation}
can be regarded as a  {\em Gaussian function of two discrete variables}.\\
Its  discrete Fourier transform  $ \mathfrak{F}[\mathfrak{g}_\tau] \!:\!\{-\!s,\!-s\!+\!1,...,s\!-\!1,s\}\!\times\!\{-\!s,\!-s\!+\!1,...,s\!-\!1,s\}\!\longrightarrow\!\mathbb{R},$
\begin{equation}
\begin{array}{lll}
\mathfrak{F}[\mathfrak{g}_\tau](k_1,k_2)\!& \!\!\!\!= & \!\!\!\!\frac{1}{d}\sum\limits_{n_1=-s}^s\sum\limits_{n_2=-s}^s {\rm e}^{-\frac{2\pi {\rm i}}{d}(k_1n_1+k_2n_2)}\, \mathfrak{g}_\tau(n_1,n_2)\\[4mm]
 \!& \!\!\!\!= & \!\!\!\!\frac{1}{\sqrt{{\rm det}\,\tau}}\sum\limits_{\beta_1,\beta_2 =-\infty }^\infty \!g_{\tau^{-1}} \!\left((k_1\!+\!\beta_1 d)\sqrt{\frac{h}{d}}, (k_2\!+\!\beta_2 d)\sqrt{\frac{h}{d}}\right)\\[4mm]
 \!& \!\!\!\!= & \!\!\!\!\sum\limits_{\beta_1,\beta_2 =-\infty }^\infty \!\mathcal{F}[g_{\tau}] \!\left((k_1\!+\!\beta_1 d)\sqrt{\frac{h}{d}}, (k_2\!+\!\beta_2 d)\sqrt{\frac{h}{d}}\right).
\end{array}
\end{equation}
 is also a Gaussian function of discrete variable, and\cite{Cotfas20}
\begin{equation}
\mathfrak{F}[\mathfrak{g}_\tau]\!=\!\frac{1}{\mbox{\scriptsize $\sqrt{{\rm det}\, \tau}$}}\, \mathfrak{g}_{\tau^{-1}}.
\end{equation}
The discrete Wigner function of \ $\mathfrak{g}_\tau $ \ is\cite{Cotfas20}
\begin{equation}\label{Wig2old}
\begin{array}{l}
\mbox{\small $\mathfrak{W}_{\mathfrak{g}_\tau}\!\!:\!\{\!-\!s,\!-s\!+\!1,\!...,\!s\!-\!1,\!s\!\}\!\!\times\!\!\{\!-\!s,\!-s\!+\!1,\!...,\!s\!-\!1,\!s\!\}\!\times\!\{\!-\!s,\!-s\!+\!1,\!...,\!s\!-\!1,\!s\!\}\!\times\!\{\!-\!s,\!-s\!+\!1,\!...,\!s\!-\!1,\!s\!\}\!\rightarrow\!\mathbb{R},$}\\[2mm]
\mbox{\small $\mathfrak{W}_{\mathfrak{g}_\tau}(n_1,n_2,k_1,k_2)\!=\!\frac{1}{d^2}\!\sum\limits_{m_1=-s}^s\sum\limits_{m_2=-s}^s\! {\rm e}^{- \frac{4\pi {\rm i}}{d}(k_1m_1+k_2m_2)}\,\mathfrak{g}_\tau(n_1\!+\!m_1,n_2\!+\!m_2)\, \overline{\mathfrak{g}_\tau(n_1\!-\!m_1,n_2\!-\!m_2)}$}\\
\qquad \qquad \qquad\quad\!=\frac{1}{\sqrt{{\rm det}\,\tau }}\ \mathfrak{g}_{2\tau}(n_1,n_2)\left[\mathfrak{g}_{2\tau^{-1}}(k_1,k_2)+\mathfrak{g}_{2\tau^{-1}}^{+0}(k_1,k_2)+
\mathfrak{g}_{2\tau^{-1}}^{0+}(k_1,k_2)+\mathfrak{g}_{2\tau^{-1}}^{++}(k_1,k_2)\right]\\[5mm]
\qquad \qquad \qquad\quad\ +\frac{1}{\sqrt{{\rm det}\,\tau}}\ \mathfrak{g}_{2\tau}^{+0}(n_1,n_2)\left[
\mathfrak{g}_{2\tau^{-1}}(k_1,k_2)-\mathfrak{g}_{2\tau^{-1}}^{+0}(k_1,k_2)+
\mathfrak{g}_{2\tau^{-1}}^{0+}(k_1,k_2)-\mathfrak{g}_{2\tau^{-1}}^{++}(k_1,k_2)\right]\\[5mm]
\qquad \qquad \qquad\quad\ +\frac{1}{\sqrt{{\rm det}\,\tau}}\ \mathfrak{g}_{2\tau}^{0+}(n_1,n_2)\left[
\mathfrak{g}_{2\tau^{-1}}(k_1,k_2)+\mathfrak{g}_{2\tau^{-1}}^{+0}(k_1,k_2)-
\mathfrak{g}_{2\tau^{-1}}^{0+}(k_1,k_2)-\mathfrak{g}_{2\tau^{-1}}^{++}(k_1,k_2)\right]\\[5mm]
 \qquad \qquad \qquad\quad\ +\frac{1}{\sqrt{{\rm det}\,\tau}}\ \mathfrak{g}_{2\tau}^{++}(n_1,n_2)\left[
\mathfrak{g}_{2\tau^{-1}}(k_1,k_2)-\mathfrak{g}_{2\tau^{-1}}^{+0}(k_1,k_2)-
\mathfrak{g}_{2\tau^{-1}}^{0+}(k_1,k_2)+\mathfrak{g}_{2\tau^{-1}}^{++}(k_1,k_2)\right],
\end{array}
\end{equation}
where
\begin{equation}\label{gsigp} 
\begin{array}{l}
\mathfrak{g}_\tau^{+0}(n_1,n_2)\!=\!\sum\limits_{\alpha_1,\alpha_2 =-\infty }^\infty {\rm e}^{-\frac{\pi}{d}(n_1\!+\!(\alpha_1 \!+\!\frac{1}{2})d\ n_2\!+\!\alpha_2 d)\, {\scriptsize \left(\!\!\begin{array}{cc}
a & \!b\\
b & \!c\end{array}\!\!\right)}\, {\scriptsize \left(\!\!\begin{array}{c}
n_1\!+\!(\alpha_1 +\frac{1}{2})d\\
n_2\!+\!\alpha_2 d\end{array}\!\!\right)}},\\[3mm] 
\mathfrak{g}_\tau^{0+}(n_1,n_2)\!=\!\sum\limits_{\alpha_1,\alpha_2 =-\infty }^\infty {\rm e}^{-\frac{\pi}{d}(n_1\!+\!\alpha_1 d\ n_2\!+\!(\alpha_2+\frac{1}{2}) d)\,  {\scriptsize \left(\!\!\begin{array}{cc}
a & \!b\\
b & \!c\end{array}\!\!\right)} \, {\scriptsize \left(\!\!\begin{array}{c}
n_1\!+\!\alpha_1 d\\
n_2\!+\!(\alpha_2\!+\!\frac{1}{2}) d\end{array}\!\!\right)}},\\[3mm] 
\mathfrak{g}_\tau^{++}(n_1,n_2)\!\!=\!\sum\limits_{\alpha_1,\alpha_2 =-\infty }^\infty {\rm e}^{-\frac{\pi}{d}(n_1\!+\!(\alpha_1 +\frac{1}{2})d\ n_2\!+\!(\alpha_2+\frac{1}{2}) d)\,  {\scriptsize \left(\!\!\begin{array}{cc}
a & \!b\\
b & \!c\end{array}\!\!\right)}\, {\scriptsize \left(\!\!\begin{array}{c}
n_1\!+\!(\alpha_1 \!+\!\frac{1}{2})d\\
n_2\!+\!(\alpha_2\!+\!\frac{1}{2}) d\end{array}\!\!\right)}}.
\end{array}
\end{equation} 
This formula, obtained by Cotfas\cite{Cotfas20}, can be written as
\begin{equation}\label{mixtwo}
\begin{array}{l}
{\small \mathfrak{W}}_{\!\mathfrak{g}_{\tau}}(n_1,n_2,k_1,k_2)\!=\!C \sum\limits_{\alpha_1,\alpha_2 =-\infty }^\infty \ 
\sum\limits_{\beta_1,\beta_2 =-\infty }^\infty (-1)^{\alpha_1\beta_1+\alpha_2\beta_2}\\[4mm]
\qquad \quad \qquad \qquad \qquad \qquad \ \ \times w\!\left(n_1\!+\!\alpha_1\frac{d}{2},n_2\!+\!\alpha_2\frac{d}{2},k_1\!+\!\beta_1\frac{d}{2},k_2\!+\!\beta_2\frac{d}{2}\right),
\end{array}
\end{equation}
where $C$ is a normalizing factor, and
\begin{equation}\label{mixtwo1}
w(q_1,q_2,p_1,p_2)\!=\!\mathrm{exp}\left\{- \mbox{\small $\frac{2\pi }{d}$}\mbox{\scriptsize  (\,$q_1$\ \ $q_2$\ \ $p_1$\ \ $p_2$\, )}{\scriptsize \left(\!\!\!
\begin{array}{cc}
\tau^{-1} & \!\!\!\!\!\!\!\!\begin{array}{cc}
0 & \!0\\
0 & \!0\end{array}\\
\begin{array}{cc}
0 & \!0\\
0 & \!0\end{array} & \!\!\!\!\!\tau
\end{array}\!\!\!\right)^{\!\!\!-1} \!{\scriptsize \left(\!\!\!
\begin{array}{c}
q_1\\
q_2\\
p_1\\
p_2
\end{array}\!\!\!\!\right)}}\right\}.
\end{equation}
\mbox{}\\
\begin{center}
 \textbf{\large III. Pure and mixed discrete variable Gaussian states}
\end{center}
{\bf Theorem 1.}\ {\it For any $\kappa \!\in\!(0,\infty)$, the pure state}
\begin{equation}
 {\bf g}_\kappa \!=\!\frac{1}{\sqrt{\langle \mathfrak{g}_\kappa,\mathfrak{g}_\kappa\rangle}}\, \mathfrak{g}_\kappa
\end{equation}
\mbox{}\qquad \qquad  \quad \mbox{\it is a discrete variable Gaussian state with $\sigma\!=\!\sigma_\kappa$, where}
\begin{equation}
 \sigma_\kappa ={\scriptsize \left(\!\!
\begin{array}{cc}
\kappa^{-1} & \!\!0\\
0 & \!\!\kappa
\end{array}\!\!\right)}.\\[3mm]
\end{equation}
{\it Proof.} Direct consequence of the  relations (\ref{gkappa}) and (\ref{gkappa1}).\quad $\rule{2mm}{2mm}$\\[5mm]
{\bf Theorem 2.}\ \mbox{\it For any $\kappa \!\in\!(0,\infty)$ and any $n_0,\, k_0\!\in\!\{-\!s,\!-s\!+\!1,...,s\!-\!1,s\}$,  the pure state}
\begin{equation}
 \psi\!=\!\mathfrak{D}(n_0,k_0){\bf g}_\kappa 
\end{equation}
\mbox{}\qquad \qquad \quad \mbox{\it is a discrete variable Gaussian state with $\sigma\!=\!\sigma_\kappa$.}\\[3mm]
{\it Proof.} We have
\begin{equation}
\psi(m)\!=\!\mathfrak{D}(n_0 ,k_0 ){\bf g}_\kappa  (m)={\rm e}^{-\frac{\pi {\rm i}}{d} n_0 k_0 }\,{\rm e}^{\frac{2\pi {\rm i}}{d}k_0m}\, {\bf g}_\kappa  (m\!-\!n_0 )
\end{equation}
and
\begin{equation}
\begin{array}{ll}
\mathfrak{W}_{\psi}(n,\!k) &\!\!\!\!=\!\frac{1}{\mbox{\footnotesize $d$}}\!\sum\limits_{m=-s}^s \!{\rm e}^{-\frac{4\pi {\rm i}}{d}km}\, {\rm e}^{\frac{2\pi {\rm i}}{d}k_0(n+m)}\, {\bf g}_\kappa  (n\!+\!m\!-\!n_0 )\, {\rm e}^{-\frac{2\pi {\rm i}}{d}k_0(n-m)}\, {\bf g}_\kappa  (n\!-\!m\!-\!n_0 )\\
&\!\!\!\!=\!\frac{1}{\mbox{\footnotesize $d$}}\!\sum\limits_{m=-s}^s \!{\rm e}^{-\frac{4\pi {\rm i}}{d}km}\, {\rm e}^{\frac{4\pi {\rm i}}{d}k_0m}\, {\bf g}_\kappa  (n\!+\!m\!-\!n_0 )\,  {\bf g}_\kappa  (n\!-\!m\!-\!n_0 )\\
&\!\!\!\!=\!\frac{1}{\mbox{\footnotesize $d$}}\!\sum\limits_{m=-s}^s \!{\rm e}^{-\frac{4\pi {\rm i}}{d}(k-k_0)m}\,  {\bf g}_\kappa  (n\!-\!n_0\!+\!m )\,  {\bf g}_\kappa  (n\!-\!n_0\!-\!m )\\
&\!\!\!\!=\!\mathfrak{W}_{{\bf g}_\kappa }(n\!-\!n_0,\!k-k_0).\quad \rule{2mm}{2mm}
\end{array}
\end{equation}
{\bf Theorem 3.}\ \mbox{\it If $\bm{\varrho}$ is a discrete variable Gaussian state, then}\\
 \mbox{}\qquad \qquad \ \mbox{\it $\mathfrak{D}(n_0,k_0)\bm{\varrho}\mathfrak{D}^\dag(n_0,k_0)$ is also a discrete variable Gaussian state,}\\
 \mbox{}\qquad \qquad  \ \mbox{\it for any $n_0,\, k_0\!\in\!\{-\!s,\!-s\!+\!1,...,s\!-\!1,s\}$.}\\[3mm]
{\it Proof.} Since $\mathfrak{D}^\dag(n_0,k_0)\Pi(n,k)\mathfrak{D}(n_0,k_0)\!=\!\Pi(n\!-\!n_0,k\!-\!k_0)$, we get
\begin{equation}
\begin{array}{ll}
\mathfrak{W}_{\mathfrak{D}(n_0,k_0)\bm{\varrho}\mathfrak{D}^\dag(n_0,k_0)}(n,\!k) &\!\!\!\!=\!\frac{1}{\mbox{\footnotesize $d$}}\, {\rm tr}(\mathfrak{D}(n_0,k_0)\bm{\varrho}\mathfrak{D}^\dag(n_0,k_0)\Pi(n,k))\\[2mm]
&\!\!\!\!=\!\frac{1}{\mbox{\footnotesize $d$}}\,{\rm tr}(\bm{\varrho}\mathfrak{D}^\dag(n_0,k_0)\Pi(n,k)\mathfrak{D}(n_0,k_0))\\[2mm]
&\!\!\!\!=\!\frac{1}{\mbox{\footnotesize $d$}}\, {\rm tr}(\bm{\varrho}\Pi(n\!-\!n_0,k\!-\!k_0))\\
&\!\!\!\!=\!\mathfrak{W}_{\bm{\varrho}}(n\!-\!n_0,k\!-\!k_0).\quad \rule{2mm}{2mm}
\end{array}
\end{equation}
{\bf Theorem 4.}\ \mbox{\it The Fourier transform $\mathfrak{F}\bm{\varrho}\mathfrak{F}^\dag$ of a discrete variable Gaussian   state $\bm{\varrho}$  is}\\
 \mbox{}\qquad \qquad \quad \  \ \mbox{\it also a discrete variable Gaussian state. If the matrix corresponding}\\
 \mbox{}\qquad \qquad \quad \  \ \mbox{\it  to $\bm{\varrho}$  is $\sigma$, then the  matrix corresponding to $\mathfrak{F}\bm{\varrho}\mathfrak{F}^\dag$  is $\Omega \sigma \Omega^T$, where}\\
{\small 
\begin{equation}
\qquad \Omega\!=\!\left(\begin{array}{rr}
0 & 1\\
-1 & 0
\end{array}\right).
\end{equation}
}
{\it Proof.} Since $\mathfrak{F}^\dag\mathfrak{D}(n,k)\mathfrak{F}\!=\!\mathfrak{D}(-k,n)$, we have
$\mathfrak{F}^\dag\Pi(n,k)\mathfrak{F}\!=\!\Pi(-k,n)$ and consequently
\begin{equation}
\begin{array}{ll}
\mathfrak{W}_{\mathfrak{F}\bm{\varrho}\mathfrak{F}^\dag}(n,\!k) &\!\!\!\!=\!\frac{1}{\mbox{\footnotesize $d$}}\, {\rm tr}(\mathfrak{F}\bm{\varrho}\mathfrak{F}^\dag\Pi(n,k))\\[2mm]
&\!\!\!\!=\!\frac{1}{\mbox{\footnotesize $d$}}\, {\rm tr}(\bm{\varrho}\mathfrak{F}^\dag\Pi(n,k)\mathfrak{F})\\[2mm]
&\!\!\!\!=\!\frac{1}{\mbox{\footnotesize $d$}}\, {\rm tr}(\bm{\varrho}\Pi(-k,n))\\
&\!\!\!\!=\!\mathfrak{W}_{\bm{\varrho}}(-k,n)\!=\!\mathfrak{W}_{\bm{\varrho}}(\Omega^{-1}(n,k)). \quad \rule{2mm}{2mm}
\end{array}
\end{equation}
The last two theorems show that  Fourier and displacement transforms preserve the nature\\ of discrete variable Gaussian states, that is, they are {\em discrete Gaussian transforms}.\\
\mbox{\bf Theorem 5.} \ \mbox{\it Each continuous variable Gaussian state $\rho_\sigma$ is the limit of a}\\ 
\mbox{}\qquad \qquad \quad \ \mbox{\it sequence of discrete variable Gaussian states  $\bm{\varrho}_\sigma$, namely}
\begin{equation}
 \rho_\sigma\!=\!\lim\limits_{d\rightarrow \infty} \bm{\varrho}_\sigma.\\[3mm]
\end{equation}
{\it Proof.}
Since $\!\!\lim\limits_{q^2+p^2\rightarrow \infty}\!\!w_\sigma(q,p)\!=\!0$, for $d$ large enough, we have
\begin{equation}
 \begin{array}{l}
w_\sigma\!\left( n\!+\!\mbox{\small{$\alpha \frac{d}{2}$}}, k\!+\! \mbox{\small{$\beta \frac{d}{2}$}}\right)\!\approx\!0 \quad \mbox{for}\ \ (\alpha,\beta)\!\not=\!(0,0),
\end{array}
\end{equation}
and consequently, up to a normalizing constant $C$,  
\begin{equation}
\begin{array}{l}
\mathfrak{W}_{\bm{\varrho}_\sigma}(n,k)\!\approx\!C\,  \mathcal{W}_{\rho_\sigma}\left(n\mbox{\small $\sqrt{\frac{h}{d}}$},k\mbox{\small $\sqrt{\frac{h}{d}}$} \right). 
\end{array}
\end{equation}
The set
\begin{equation}
\begin{array}{l}
\left\{ \left. \ \left(n\mbox{\small $\sqrt{\frac{h}{d}}$},k\mbox{\small $\sqrt{\frac{h}{d}}$} \right)\ \ \right|\ \ 
\begin{array}{l}
d\!=\!2s\!+\!1\!\in \!\{ 3,5,7,...\}\\
n,k\!\in\!\{-s,-s\!+\!1,...,s\!-\!1,s\}
\end{array}\ \right\}
\end{array}
\end{equation}
being dense in the phase space $\mathbb{R}^2$, the function $\mathcal{W}_{\rho_\sigma}\!:\!\mathbb{R}\!\times\!\mathbb{R}\!\rightarrow\!\mathbb{R}$ is determined by the functions\\
$\mathfrak{W}_{\bm{\varrho}_\sigma}\!:\!\{-\!s,\!-s\!+\!1,...,s\!-\!1,s\}\!\times\!\{-\!s,\!-s\!+\!1,...,s\!-\!1,s\}\!\longrightarrow\!\mathbb{R}.$ corresponding to $d\!\in \!\{ 3,5,7,...\}.\ \ \rule{2mm}{2mm}$\\
\mbox{\bf Theorem 6.} \ \mbox {\it For any covariance matrix $\sigma$, we have}
\begin{equation}
\lim\limits_{d\rightarrow \infty}{\rm tr}\, \bm{\varrho}_{\sigma}^2\!=\!\frac{1}{\sqrt{\det \sigma}}.\\[3mm]
\end{equation}
{\it Proof}. For $d$ large enough, we have $\mathfrak{W}_{\bm{\varrho}_\sigma}(n,k)\!\approx\!C\,  \mathcal{W}_{\rho_\sigma}\left(n\mbox{\small $\sqrt{\frac{h}{d}}$},k\mbox{\small $\sqrt{\frac{h}{d}}$} \right)$, and consequently
\begin{equation}
{\rm tr}\, \bm{\varrho}_{\sigma}^2\!=\!d\!\!\sum_{n,k=-s}^s\!\! \mathfrak{W}_{\bm{\varrho}_{\sigma}}^2(n,k)\!\approx\!d
\frac{\sum\limits_{n,k=-s}^s\mathcal{W}^2_{\rho_\sigma}\left(n\mbox{\small $\sqrt{\frac{h}{d}}$},k\mbox{\small $\sqrt{\frac{h}{d}}$} \right)}
{\left(\sum\limits_{n,k=-s}^s \mathcal{W}_{\rho_\sigma}\left(n\mbox{\small $\sqrt{\frac{h}{d}}$},k\mbox{\small $\sqrt{\frac{h}{d}}$} \right) \right)^2}.
\end{equation}
By considering a partition of the rectangle $\left[-\frac{d}{2}\mbox{\small $\sqrt{\frac{h}{d}}$},\frac{d}{2}\mbox{\small $\sqrt{\frac{h}{d}}$}\right]\!\times\!\left[-\frac{d}{2}\mbox{\small $\sqrt{\frac{h}{d}}$},\frac{d}{2}\mbox{\small $\sqrt{\frac{h}{d}}$}\right]$
into $d^2$ squares of area $\frac{h}{d}$\\ and regarding the integrals as limits of Riemann sums, we get
\begin{equation}
\begin{array}{l}
{\rm tr}\, \bm{\varrho}_{\sigma}^2\!\approx\! h 
\frac{\sum\limits_{n,k=-s}^s\frac{h}{d}\, \mathcal{W}^2_{\rho_\sigma}\left(n\mbox{\small $\sqrt{\frac{h}{d}}$},k\mbox{\small $\sqrt{\frac{h}{d}}$} \right)}
{\left(\sum\limits_{n,k=-s}^s\frac{h}{d}\,\mathcal{W}_{\rho_\sigma}\left(n\mbox{\small $\sqrt{\frac{h}{d}}$},k\mbox{\small $\sqrt{\frac{h}{d}}$}\right) \right)^2}  \stackrel{d\rightarrow \infty}{-\!\!\!-\!\!\!\longrightarrow} 
\frac{h \int\limits_{\mathbb{R}^2}\mathcal{W}^2_{\rho_\sigma}(q,p)\, dq\, dp} 
{\left(\int\limits_{\mathbb{R}^2} \mathcal{W}_{\rho_\sigma}(q,p)\, dq\, dp\right)^2} \!=\!{\rm tr}\, \rho_{\sigma}^2
\!=\! \frac{1}{\sqrt{\det \sigma}}. \quad \rule{2mm}{2mm}
\end{array}
\end{equation}
The Table I contains some numerical data obtained in certain particular cases.\\
\noindent\mbox{\bf Theorem 7.}\ \mbox{\it If ${\rm det}\, \sigma \!=\!1$, then 
the discrete Gaussian state $\bm{\varrho}_{\sigma}$ is a pure state:}
\begin{equation}
{\rm det}\, \sigma \!=\!1\quad \Rightarrow\quad   {\rm tr}\, \bm{\varrho}_{\sigma}^2\!=\!1.\\[5mm]
\end{equation}
{\it Proof.} If ${\rm det}\, \sigma \!=\!1$, then there exist $\kappa \!\in\!(0,\infty)$ and a canonical transformation (rotation) 
\begin{equation}
\left(
\begin{array}{c}
q'\\
p'
\end{array}\right)\!=\!
\left(
\begin{array}{rr}
\cos \theta & -\sin \theta\\
\sin \theta & \cos \theta 
\end{array}\right)
\left(
\begin{array}{c}
q\\
p
\end{array}\right)
\end{equation}
of the phase space such that 
\begin{equation}
\sigma \!=\!
\left(
\begin{array}{rr}
\cos \theta & \sin \theta\\
-\sin \theta & \cos \theta 
\end{array}\right)
\left(\!\!
\begin{array}{cc}
\kappa^{-1} & 0\\
0 & \kappa
\end{array}\!\!\right)
\left(
\begin{array}{rr}
\cos \theta & -\sin \theta\\
\sin \theta & \cos \theta 
\end{array}\right).
\end{equation}

Consequently
\begin{equation}
\mathcal{W}_{\rho_\sigma}(q,p)\!=\!
\mbox{\small $\frac{2}{ h}$}{\rm e}^{-\frac{2\pi }{h}\left( \kappa {q'}^2+\kappa^{-1}{p'}^2\right)}
\!=\!\mathcal{W}_{\rho_{\sigma_\kappa}}(q',p')
\end{equation}
but, we know that the discrete Wigner function $\mathfrak{W}_{\varrho_{\sigma_\kappa}}$ corresponding to $\mathcal{W}_{\rho_{\sigma_\kappa}}$ represents a pure state.\quad $\rule{2mm}{2mm}$\\
The Table II contains some numerical data obtained in certain particular cases.
\begin{center}
 \textbf{\large IV. Position-momentum commutation relation and the discrete variable Gaussian states}
\end{center}
In the continuous case, the usual position-momentum commutation relation
\begin{equation}
[\hat{q},\hat{p}]={\rm i}\mbox{\small $\frac{h}{2\pi }$}
\end{equation}
is satisfied in a subspace dense in $L^2(\mathbb{R})$. The discrete counterpart
\begin{equation}
[\hat{\mathfrak{q}},\hat{\mathfrak{p}}]={\rm i}\mbox{\small $\frac{d}{2\pi }$}
\end{equation}
is not satisfied, but for $d$ large enough, most of the eigenvalues of the operator $[\hat{\mathfrak{q}},\hat{\mathfrak{p}}]-{\rm i}\frac{d}{2\pi }$ are almost null. 


Let $\lambda _1$, $\lambda _2$, ..., $\lambda _d$ be the  eigenvalues of $[\hat{\mathfrak{q}},\hat{\mathfrak{p}}]-{\rm i}\frac{d}{2\pi }$, considered in the increasing order of their modulus ($|\lambda _1|\leq |\lambda _2|\leq ...\leq |\lambda _d|$), and let
$\varphi_1$, $\varphi_2$, ... , $\varphi_d$  be the corresponding eigenfunctions.
If we expand a discrete variable Gaussian state in terms of the eigenbasis $\{ \varphi_k\}$, the most significant coefficients seem to be those corresponding to the functions $\varphi_k$ with small $|\lambda_k|$.
For example, in the case $d\!=\!11$,   the matrix $(|\langle\varphi_n|{\bf g}_1\rangle|)$ of ${\bf g}_1$ in the  eigenbasis $\{\varphi_k\}$ is
{\scriptsize
\begin{equation}\qquad \qquad \qquad 
\left(
\begin{array}{c}
0.9999\\
1\!*\!10^{-10}\\
2\!*\!10^{-11}\\
2\!*\!10^{-13}\\
0.0079\\
4\!*\!10^{-15}\\
1\!*\!10^{-15}\\
2\!*\!10^{-16}\\
0.0004\\
6\!*\!10^{-18}\\
2\!*\!10^{-17}\\
\end{array}\right)
\end{equation}}
\noindent and for {\scriptsize $\sigma\!=\!\left(\!\!
\begin{array}{cc}
2 & \!\!1\\
1 & \!\!1
\end{array}\!\!\right)$}, the matrix $\left(\left|\left\langle \varphi_n\left|\bm{\varrho}_{\sigma}\right|\varphi_m\right\rangle\right|\right)$ of $\bm{\varrho}_{\sigma}$  is
{\scriptsize
\begin{equation}
\!\!\!\!\!\left(
\begin{array}{ccccccccccc}
0.8954 & 6\!*\!10^{-11} & 0.2837 & 1\!*\!10^{-12} & 0.1059 & 
  1\!*\!10^{-14} & 0.0393 & 3\!*\!10^{-16} & 0.0164 & 3\!*\!10^{-17} &
   0.0084\\
1\!*\!10^{-10} & 5\!*\!10^{-18} & 5\!*\!10^{-11} &  1\!*\!10^{-17} &  
  1\!*\!10^{-11} & 8\!*\!10^{-18} &  7\!*\!10^{-12} &  1\!*\!10^{-17} &  
  3\!*\!10^{-12} &  1\!*\!10^{-17} &  1\!*\!10^{-12}\\
 0.2837 & 2\!*\!10^{-11} &  0.0899 &  3\!*\!10^{-13} &  0.0335 &  
  4\!*\!10^{-15} &  0.0124 &  1\!*\!10^{-16} &  0.0052 &  
  1\!*\!10^-17 &  0.0026\\
 2\!*\!10^{-13} &  1\!*\!10^{-17} &  7\!*\!10^{-14} &  3\!*\!10^{-18} &  
  2\!*\!10^{-14} &  3\!*\!10^{-17} &  1\!*\!10^{-14} &  4\!*\!10^{-17} &  
  4\!*\!10^{-15} &  1\!*\!10^{-17} &  2\!*\!10^{-15}\\
 0.1059 & 7\!*\!10^{-12} &  0.0335 &  1\!*\!10^{-13} &  0.0125 &  
  1\!*\!10^{-15} &  0.0046 &  4\!*\!10^{-17} &  0.0019 & 
  3\!*\!10^{-18} &  0.0010\\
 1\!*\!10^{-15} &  1\!*\!10^{-17} &  5\!*\!10^{-16} &  2\!*\!10^{-17} & 
  2\!*\!10^{-16} &  1\!*\!10^{-17} &  8\!*\!10^{-17} &  1\!*\!10^{-17} &  
  3\!*\!10^{-17} &  1\!*\!10^{-17} &  7\!*\!10^{-18}\\
 0.0393 &  2\!*\!10^{-12} &  0.0124 &  5\!*\!10^{-14} &  0.0046 &  
  5\!*\!10^{-16} &  0.0017 &  2\!*\!10^{-17} &  0.0007 &  
  6\!*\!10^{-18} &  0.0003\\
 2\!*\!10^{-16} &  1\!*\!10^{-17} &  8\!*\!10^{-17} &  4\!*\!10^{-17} &  
  3\!*\!10^{-17} &  7\!*\!10^{-18} &  2\!*\!10^{-17} &  4\!*\!10^{-18} &  
  8\!*\!10^{-18} &  3\!*\!10^{-17} &  5\!*\!10^{-18}\\
 0.0164 &  1\!*\!10^{-12} &  0.0052 &   2\!*\!10^{-14} &   0.0019 &   
  2\!*\!10^{-16} &   0.0007 &   1\!*\!10^{-17} &   0.0003 &   
  8\!*\!10^{-19} &   0.0001\\
 2\!*\!10^{-17} &   1\!*\!10^{-17} &   8\!*\!10^{-18} &   1\!*\!10^{-17} &   
  2\!*\!10^{-18} &   1\!*\!10^{-17} &   5\!*\!10^{-18} &   3\!*\!10^{-17} &   
  1\!*\!10^{-18} &  4\!*\!10^{-17} &   5\!*\!10^{-18}\\
 0.0084 &   6\!*\!10^{-13} &   0.0026 &   1\!*\!10^{-14} &   0.0010 &   
  1\!*\!10^{-16} &   0.00037 &   2\!*\!10^{-18} &   0.0001 &  
  5\!*\!10^{-18} &   0.0000 
\end{array}\right).
\end{equation}} 

For $\varepsilon \!>\!0$, let $d_\varepsilon$ be such that 
 \begin{equation}
\{ \lambda _1,\, \lambda _2,\, ...\, ,\, \lambda _{d_\varepsilon}\}\!=\!\{\, \lambda_k\ |\ \ |\lambda_k|\!<\!\varepsilon\,\},
\end{equation}
and let 
\begin{equation}
\mathcal{H}_\varepsilon\!=\!{\rm span}\{\,\varphi_1,\, \varphi_2,\, ...\, ,\, \varphi_{d_\varepsilon}\}. 
\end{equation}
If $\varepsilon$ is small enough, then we can consider that 
\begin{equation}
[\hat{\mathfrak{q}},\hat{\mathfrak{p}}]\approx{\rm i}\mbox{\small $\frac{d}{2\pi }$}\quad \mbox{in}\quad \mathcal{H}_\varepsilon.
\end{equation}
By following the analogy with the continuous case, we define
\begin{equation}
\begin{array}{r}
\hat{\mathfrak{a}}\!=\!\sqrt{\frac{\pi}{d}}(\hat{\mathfrak{q}}\!+\!{\rm i}\hat{\mathfrak{p}}),\\[2mm]
\hat{\mathfrak{a}}^\dag\!=\!\sqrt{\frac{\pi}{d}}(\hat{\mathfrak{q}}\!-\!{\rm i}\hat{\mathfrak{p}}),
\end{array}\quad \mbox{similar to}\quad 
\begin{array}{r}
\hat{a}\!=\!\sqrt{\frac{\pi}{h}}(\hat{q}\!+\!{\rm i}\hat{p}),\\[2mm]
\hat{a}^\dag\!=\!\sqrt{\frac{\pi}{h}}(\hat{q}\!-\!{\rm i}\hat{p}),
\end{array}
\end{equation}
The discrete variable Gaussian states seem to mainly belong  to $\mathcal{H}_\varepsilon$ and $\mathcal{A}(\mathcal{H}_\varepsilon)$.

In the case of continuous variable, the transformations 
$\hat U\!=\! {\rm e}^{-\frac{\rm i}{2}\hat{H}}$, where
\begin{equation}
\begin{array}{l}
\hat{H}\!=\! (\,\hat{a}^\dag \ \ \hat{a}\,){\small \left(\!\!
\begin{array}{cc}
A & \!\!B\\
\bar{B} &\!\! A
\end{array}\!\!\right)}
\left(\!\!
\begin{array}{c}
\hat{a}\\
\hat{a}^\dag
\end{array}\!\!\right)\quad \mbox{with}\ \ 
\mbox{\small $\begin{array}{r}
A\!\in\!\mathbb{R},\\
B\!\in\!\mathbb{C},
\end{array}$}
\end{array}
\end{equation}
are Gaussian transforms, that is they preserve the Gaussian nature of the states.\\
For $\hat U\!=\!{\rm e}^{-\frac{\rm i}{2}\hat{H}}$,   there exists 
  $S\!\in\!{\rm Sp}(2,\mathbb{R})$ such that
\begin{equation}
{\rm e}^{-\frac{\rm i}{2}\hat{H}}\rho_\sigma {\rm e}^{\frac{\rm i}{2}\hat{H}} =\rho_{S\sigma S^T}.
\end{equation}

In the case of discrete variable, the relation $\lim\limits_{d\rightarrow \infty}\bm{\varrho}_\sigma=\rho_\sigma$ suggests that for
\begin{equation}
\begin{array}{l}
\hat{\mathfrak{H}}\!=\! (\,\hat{\mathfrak{a}}^\dag \ \ \hat{\mathfrak{a}}\,){\small \left(\!\!
\begin{array}{cc}
A & \!\!B\\
\bar{B} &\!\! A
\end{array}\!\!\right)}
\left(\!\!
\begin{array}{c}
\hat{\mathfrak{a}}\\
\hat{\mathfrak{a}}^\dag
\end{array}\!\!\right)\quad \mbox{with}\ \ 
\mbox{\small $\begin{array}{l}
A\!\in\!\mathbb{R},\\
B\!\in\!\mathbb{C},
\end{array}$}
\end{array}
\end{equation}
we must have
\begin{equation}\label{discGaussTr}
{\rm e}^{-\frac{\rm i}{2}\hat{\mathfrak{H}}}\bm{\varrho}_\sigma {\rm e}^{\frac{\rm i}{2}\hat{\mathfrak{H}}} \approx\bm{\varrho}_{S\sigma S^T}\qquad \mbox{for}\ \ d\ \ \mbox{large enough}.
\end{equation}
Some numerical results concerning a phase shift and a single-mode squeezing transformation are presented in Table III.\\
A possible explanation can be obtained by following the analogy with the continuous case. From the relation
\begin{equation}
[\hat{\mathfrak{a}},\hat{\mathfrak{a}}^\dag]\approx 1
\end{equation}
satisfied in $\mathcal{H}_\varepsilon$, one gets the relations
\begin{equation}
\begin{array}{l}
\left[\frac{\rm i}{2}\hat{\mathfrak{H}}, 
\hat{\mathfrak{a}}\right]\!\approx \!{\rm i}
\left(
-A \, \hat{\mathfrak{a}}\!-\!B\,\hat{\mathfrak{a}}^\dag \right)\\[2mm]
\left[\frac{\rm i}{2}\hat{\mathfrak{H}} , 
\hat{\mathfrak{a}}^\dag\right]\!\approx \!{\rm i}
\left(
\bar{B} \, \hat{\mathfrak{a}}\!+\!A\,\hat{\mathfrak{a}}^\dag \right)
\end{array}\qquad \mbox{in}\quad \mathcal{H}_\varepsilon
\end{equation}
which can be written together as
\begin{equation}
\mbox{\small $\left[\frac{\rm i}{2}\hat{\mathfrak{H}}, \left(\!\!
\begin{array}{c}
\hat{\mathfrak{a}}\\
\hat{\mathfrak{a}}^\dag
\end{array}\!\!\right)\right]\!\approx \!{\rm i}
\mbox{\small $\left(\!\!\!
\begin{array}{rr}
-A & -B\\
\bar{B} & \!A
\end{array}\!\!\!\right)$}
\left(\!\!
\begin{array}{c}
\hat{\mathfrak{a}}\\
\hat{\mathfrak{a}}^\dag
\end{array}\!\!\right)$}\qquad \mbox{in}\quad \mathcal{H}_\varepsilon.
\end{equation}
By iteration, we obtain
\begin{equation}
\mbox{\small $\left[\frac{\rm i}{2}\hat{\mathfrak{H}}, 
\left[\frac{\rm i}{2}\hat{\mathfrak{H}}, \left(\!\!
\begin{array}{c}
\hat{\mathfrak{a}}\\
\hat{\mathfrak{a}}^\dag
\end{array}\!\!\right)\right]\right]\!\approx \!{\rm i}^2
\mbox{\small $\left(\!\!\!
\begin{array}{rr}
-A & -B\\
\bar{B} & \!A
\end{array}\!\!\!\right)$}^2
\left(\!\!
\begin{array}{c}
\hat{\mathfrak{a}}\\
\hat{\mathfrak{a}}^\dag
\end{array}\!\!\right)$}\qquad \mbox{in}\quad \mathcal{H}_\varepsilon,
\end{equation}
\begin{equation} 
\mbox{\small $\left[\frac{\rm i}{2}\hat{\mathfrak{H}}, 
\left[\frac{\rm i}{2}\hat{\mathfrak{H}}, 
\left[\frac{\rm i}{2}\hat{\mathfrak{H}}, \left(\!\!
\begin{array}{c}
\hat{\mathfrak{a}}\\
\hat{\mathfrak{a}}^\dag
\end{array}\!\!\right)\right]\right]\right]\!\approx \!{\rm i}^3
\mbox{\small $\left(\!\!\!
\begin{array}{rr}
-A & -B\\
\bar{B} & \!A
\end{array}\!\!\!\right)$}^3
\left(\!\!
\begin{array}{c}
\hat{\mathfrak{a}}\\
\hat{\mathfrak{a}}^\dag
\end{array}\!\!\right)$},\ \mbox{etc.}\qquad \mbox{in}\quad\mathcal{H}_\varepsilon.
\end{equation}
From the relation
\begin{equation} 
{\rm e}^{X}Y{\rm e}^{-X}=Y+\frac{1}{1!}[X,Y]+\frac{1}{2!}[X,[X,Y]]+
\frac{1}{3!}[X,[X,[X,Y]]]+ \cdots
\end{equation}
it follows that $\hat{\mathfrak{U}}\!=\!{\rm e}^{-\frac{\rm i}{2}\hat{\mathfrak{H}}}$ satisfies the relation
\begin{equation}
\mbox{\small $\hat{\mathfrak{U}}^\dag\left(\!\!
\begin{array}{c}
\hat{\mathfrak{a}}\\
\hat{\mathfrak{a}}^\dag
\end{array}\!\!\right)
\hat{\mathfrak{U}} \approx  {\rm exp}\left\{{\rm i}
\mbox{\small $\left(\!\!\!
\begin{array}{rr}
-A & -B\\
\bar{B} & \!A
\end{array}\!\!\!\right)$}\right\}\left(\!\!
\begin{array}{c}
\hat{\mathfrak{a}}\\
\hat{\mathfrak{a}}^\dag
\end{array}\!\!\right)$}\qquad \mbox{in}\quad \mathcal{H}_\varepsilon.
\end{equation}

Since\\[-5mm]
\begin{equation}
\mbox{\small $\left(\!\!
\begin{array}{c}
\hat{\mathfrak{a}}\\
\hat{\mathfrak{a}}^\dag
\end{array}\!\!\right)\!=\!\sqrt{\frac{\pi }{d}}
\left(\!\!
\begin{array}{rr}
1 & {\rm i}\\
1 & -{\rm i}
\end{array}\!\!\right)
\left(\!\!
\begin{array}{c}
\hat{\mathfrak{q}}\\
\hat{\mathfrak{p}}
\end{array}\!\!\right)$},
\end{equation}
the previous relation can be written as
\begin{equation}
\mbox{\small $\hat{\mathfrak{U}}^\dag\left(\!\!
\begin{array}{c}
\hat{\mathfrak{q}}\\
\hat{\mathfrak{p}}
\end{array}\!\!\right)
\hat{\mathfrak{U}} \approx  S\left(\!\!
\begin{array}{c}
\hat{\mathfrak{q}}\\
\hat{\mathfrak{p}}
\end{array}\!\!\right)$}\qquad \mbox{in}\quad \mathcal{H}_\varepsilon,\\[-5mm]
\end{equation}
where\\[-5mm]
\begin{equation} 
\mbox{\small $\begin{array}{l}
S\!=\!\frac{1}{2}\left(\!\!
\begin{array}{rr}
1 & 1\\
-{\rm i} & {\rm i}
\end{array}\!\!\right){\rm exp}\left\{{\rm i}
\mbox{\small $\left(\!\!\!
\begin{array}{rr}
-A & -B\\
\bar{B} & \!A
\end{array}\!\!\!\right)$}\right\}
\left(\!\!
\begin{array}{rr}
1 & {\rm i}\\
1 & -{\rm i}
\end{array}\!\!\right)\!=\!{\rm exp}\left\{-\frac{1}{2}
\left(\!\!
\begin{array}{cc}
(B-\bar{B}){\rm i} & -2A+B+\bar{B}\\
2A+B+\bar{B} & -(B-\bar{B}){\rm i}
\end{array}\!\!\right)\right\}.
\end{array}$}
\end{equation}
This matrix with real elements and $\det S\!=\!1$ belongs to  the symplectic group ${\rm Sp}(2,\mathbb{R})$.\\
For example,  the phase shift $U\!=\!{\rm e}^{\frac{\rm i}{2}(\varphi  \, \hat{a}^\dag \hat{a}+ \varphi  \, \hat{a} \hat{a}^\dag)}$ corresponds to
\begin{equation}
S_\varphi\!=\!\left(\!\!
\begin{array}{cc}
\cos \varphi & -\sin \varphi\\[1mm]
\sin \varphi & \cos \varphi
\end{array}\!\!\right)
\end{equation}
and the single-mode squeezing $U\!=\!{\rm e}^{\frac{s}{2}\left({\rm e}^{{\rm i}\theta}(\hat{a}^\dag)^2-{\rm e}^{-{\rm i}\theta} \, \hat{a}^2\right)}$ to
\begin{equation}
S_{s,\theta}\!=\!\left(\!\!
\begin{array}{cc}
\cosh s\!+\!\cos \theta \ \sinh s & \sin \theta \, \sinh s\\
\sin \theta \, \sinh s & \cosh s\!-\!\cos \theta \ \sinh s
\end{array}\!\!\right) . 
\end{equation}

In $\mathcal{H}_\varepsilon$, we have $[\hat{\mathfrak{q}},[\hat{\mathfrak{q}},\hat{\mathfrak{p}}]]\approx 0$ and $[\hat{\mathfrak{p}},[\hat{\mathfrak{q}},\hat{\mathfrak{p}}]]\approx 0$. Consequently,
\begin{equation}
\begin{array}{l}
\mathfrak{D}(n ,k )\!=\!{\rm e}^{-\frac{\pi {\rm i}}{d} n k }\,{\rm e}^{\frac{2\pi {\rm i}}{d}k \hat{\mathfrak{q}}}\, {\rm e}^{-\frac{2\pi {\rm i}}{d}n \hat{\mathfrak{p}}}\\
\qquad \quad \ \!=\!{\rm e}^{-\frac{\pi {\rm i}}{d} n k }\,{\rm e}^{\frac{1}{2}\left[\frac{2\pi {\rm i}}{d}k \hat{\mathfrak{q}},-\frac{2\pi {\rm i}}{d}n \hat{\mathfrak{p}}\right] }\, {\rm e}^{-\frac{2\pi {\rm i}}{d}(n \hat{\mathfrak{p}}-k \hat{\mathfrak{q}})}\\
\qquad \quad \ \!\approx \! {\rm e}^{-\frac{2\pi {\rm i}}{d}(n \hat{\mathfrak{p}}-k \hat{\mathfrak{q}})}
\end{array}
\end{equation}
and by denoting {\scriptsize $\mbox{{\normalsize $S$}}\!=\!\left( \!\!\begin{array}{cc}
s_{11} & s_{12}\\
s_{21} & s_{22}
\end{array}\!\!
\right)$}, we get
\begin{equation}
\begin{array}{l}
\hat{\mathfrak{U}}^\dag\mathfrak{D}(n ,k )\hat{\mathfrak{U}} \approx  {\rm e}^{-\frac{2\pi {\rm i}}{d}\hat{\mathfrak{U}}^\dag(n \hat{\mathfrak{p}}-k \hat{\mathfrak{q}})\hat{\mathfrak{U}} } \\
\qquad \quad \qquad =\!{\rm e}^{-\frac{2\pi {\rm i}}{d}(n (s_{21}\hat{\mathfrak{q}}+s_{22}\hat{\mathfrak{p}})-k (s_{11}\hat{\mathfrak{q}}+s_{12}\hat{\mathfrak{p}}))} \\
\qquad \quad \qquad =\!{\rm e}^{-\frac{2\pi {\rm i}}{d}( (s_{22}n-s_{12}k)\hat{\mathfrak{p}}- (-s_{21}n+s_{11}k)\hat{\mathfrak{q}}))} \\
\qquad \quad \qquad \approx  \mathfrak{D}(S^{-1}(n ,k) ).
\end{array}
\end{equation}
Since 
\begin{equation} 
\begin{array}{l}
w_\sigma(S^{-1}(q,p))\!=\!{\rm exp}\left\{- \mbox{\small  (\,$q$\ \ $p$)}(S^{-1})^T\sigma^{\!-1} S^{-1}\!\left(\!\!\!
\begin{array}{c}
q\\
p
\end{array}\!\!\!\!\right)\right\}\\
\qquad \qquad \qquad =\!{\rm exp}\left\{- \mbox{\small  (\,$q$\ \ $p$)}(S\sigma S^T)^{-1}\!\left(\!\!\!
\begin{array}{c}
q\\
p
\end{array}\!\!\!\!\right)\right\}
\!=\!w_{S\sigma S^T}(q,p)
\end{array}
\end{equation}
and 
\begin{equation}
\begin{array}{lllll}
\Pi^2\!=\!\mathbb{I},\qquad & \Pi \mathfrak{F}\!=\!\mathfrak{F}\Pi, \qquad & \Pi \hat{\mathfrak{q}}\!=\!-\hat{\mathfrak{q}}\Pi,\qquad  & \Pi \hat{\mathfrak{a}}\!=\!-\hat{\mathfrak{a}}\Pi,\qquad  & \Pi \hat{\mathfrak{H}}\Pi \!=\!\hat{\mathfrak{H}}.\\
\mathfrak{F}^2\!=\!\Pi, \qquad & \Pi \mathfrak{F}^\dag \!=\!\mathfrak{F}^\dag \Pi, \qquad & \Pi \hat{\mathfrak{p}}\!=\!-\hat{\mathfrak{p}}\Pi,\qquad  & \Pi \hat{\mathfrak{a}}^\dag\!=\!-\hat{\mathfrak{a}}^\dag\Pi,\qquad  & \Pi \hat{\mathfrak{U}}\Pi \!=\!\hat{\mathfrak{U}},
\end{array}
\end{equation}
we get $\hat{\mathfrak{U}}^\dag \Pi \hat{\mathfrak{U}}\!=\!\Pi$ and 
\begin{equation}
\begin{array}{l}
\mathfrak{W}_{\hat{\mathfrak{U}}\bm{\varrho}_\sigma \hat{\mathfrak{U}}^\dag }(n,k)\!=\!\frac{1}{d}\ 
{\rm tr}(\hat{\mathfrak{U}}\bm{\varrho}_\sigma \hat{\mathfrak{U}}^\dag \,\Pi (n,k))\!=\!\frac{1}{d}\ {\rm tr}(\bm{\varrho}_\sigma \hat{\mathfrak{U}}^\dag \,
\Pi (n,k)\,\hat{\mathfrak{U}})\\
\qquad \qquad \quad \ \ =\!\frac{1}{d}\ {\rm tr}(\bm{\varrho}_\sigma \hat{\mathfrak{U}}^\dag \,\mathfrak{D}(n ,k )\, \Pi \,\mathfrak{D}^\dag(n ,k )\,\hat{\mathfrak{U}})\\
\qquad \qquad \quad \ \ =\!\frac{1}{d}\ {\rm tr}(\bm{\varrho}_\sigma \hat{\mathfrak{U}}^\dag \,\mathfrak{D}(n ,k )\, \hat{\mathfrak{U}}\,\hat{\mathfrak{U}}^\dag \, 
\Pi \, \hat{\mathfrak{U}}\,\hat{\mathfrak{U}}^\dag \, \mathfrak{D}^\dag(n ,k )\,\hat{\mathfrak{U}})\\
\qquad \qquad \quad \ \ \approx\!\frac{1}{d}\ {\rm tr}(\bm{\varrho}_\sigma \,\mathfrak{D}(S^{-1}(n ,k))\, \Pi \, 
\mathfrak{D}^\dag(S^{-1}(n ,k) ))\\
\qquad \qquad \quad \ \ =\!\frac{1}{d}\ {\rm tr}(\bm{\varrho}_\sigma \,\Pi (S^{-1}(n ,k))\!=\!
\mathfrak{W}_{\bm{\varrho}_\sigma }(S^{-1}(n,k))\\
\qquad \qquad \quad \ \ =\!C\!\sum\limits_{\alpha,\beta  =-\infty }^\infty \!(-1)^{\alpha \beta}\ 
w_\sigma\!\left( S^{-1}(n,k)\!+\!(\alpha ,\beta) \frac{d}{2}\right).\\
\end{array}
\end{equation}
If $d$ is large enough, then \ \ 
$
w_\sigma\!\left( S^{-1}(n,k)\!+\!(\alpha ,\beta) \frac{d}{2}\right)\!\approx \! 0 \quad \mbox{for}\quad (\alpha ,\beta)\neq (0,0),\ \ 
$
and we get the relation
\begin{equation}
\begin{array}{l}
\mathfrak{W}_{\hat{\mathfrak{U}}\bm{\varrho}_\sigma \hat{\mathfrak{U}}^\dag }(n,k)\!\approx\!C\ w_\sigma\!\left( S^{-1}(n,k)\right)\\
\qquad \qquad \quad \ \ =\!C\ w_{S\sigma S^T}\!\left(n,k\right)\\
\qquad \qquad \quad \ \ \approx\!\mathfrak{W}_{\bm{\varrho}_{S\sigma S^T}}(n,k)
\end{array}
\end{equation}
equivalent to
\begin{equation}
\hat{\mathfrak{U}}\bm{\varrho}_\sigma \hat{\mathfrak{U}}^\dag \!\approx\!\bm{\varrho}_{S\sigma S^T}\qquad \mbox{in}\quad \mathcal{H}_{\varepsilon}.
\end{equation}
---------------------------------\\
More details and the used source codes can be found in my note
"PURE AND MIXED DISCRETE VARIABLE GAUSSIAN STATES.
ARE THEY SOME NEW SIGNIFICANT QUANTUM STATES ?" available on my website \mbox{https://unibuc.ro/user/nicolae.cotfas/},\ \  at the section "Documente". 
\end{widetext}
\end{document}